\documentstyle[12pt,aps,eqsecnum,amsfonts,amssymb]{revtex}
\tighten
\newcommand{\be}{\begin{equation}}
\newcommand{\ee}{\end{equation}}
\newcommand{\ba}{\begin{array}}
\newcommand{\ea}{\end{array}}
\newcommand{\bea}{\begin{eqnarray}}
\newcommand{\eea}{\end{eqnarray}}
\newcommand*{\ccc}[1]{\stackrel{*}{#1}\!\!{\vphantom{#1}}}
\newcommand*{\cc}[1]{ \rlap{$\stackrel{*}{\phantom{#1}}$}#1 }
\newcommand*{\oo}[1]{ \rlap{$\stackrel{\circ}{\phantom{#1}}$}#1 }
\newcommand{\uGamma}{\hat{\underline\Gamma}{\vphantom{\Gamma}}}
\newcommand{\uz}{{\underline z}{\vphantom{z}}}
\newcommand{\bz}{{\bf z}}
\newcommand{\qq}{\frac 12}
\newcommand{\diag}{\mathop{{\rm diag}}}

\newcommand{\sign}{\mathop{{\rm sign}}}
\newcommand{\da}{{\dot\alpha}}
\newcommand{\db}{{\dot\beta}}

\newcommand{\eps}{\varepsilon}

\begin{document}
%%%%%%%%%%%%%%%%%%%%%%%%%%%%%%%%%%%%%%%%%%%%%%%%%%%%%%%%%%%%%%%%%%%%%%%%%%%%%
\title{DISCRETE SYMMERTRIES AS AUTOMORPHISMS OF THE PROPER POINCAR\'E GROUP}
\author{I L Buchbinder${}^{a,b}$,
        D M Gitman${}^a$,
        A L Shelepin${}^{a,c}$
\thanks{E-mail: joseph@tspu.edu.ru, gitman@fma.if.usp.br, alex@shelepin.msk.ru}
       }
\address {${}^a$Instituto de F{\'\i}sica, Universidade de S\~ao Paulo,
          Caixa Postal 66318, 05315-970--S\~ao Paulo, SP, Brazil \\
          ${}^b${Department of Theoretical Physics, Tomsk
          State Pedagogical University, Tomsk 634041, Russia}   \\
          ${}^c${Moscow Institute of Radio Engenering, Electronics
          and Automation, Prospect Vernadskogo, 78, 117454, Moscow, Russia} }
\maketitle

\bigskip

{\it
We present the consistent approach to finding the discrete transformations
in the representation spaces of the proper Poincar\'e group. To this end we
use the possibility to establish a correspondence between involutory
automorphisms of the proper Poincar\'e group and the discrete transformations.
As a result, we derive rules of the discrete transformations for arbitrary
spin-tensor fields without the use of relativistic wave equations. Besides,
we construct explicitly fields carrying representations of the extended
Poincar\'e group, which includes the discrete transformations as well.
}

%%%%%%%%%%%%%%%%%%%%%%%%%%%%%%%%%%%%%%%%%%%%%%%%%%%%%%%%%%%%%%%%%%%%%%%%%%%%%
\section{Introduction}

As it is well known, the Lorentz transformations in Minkowski space
are divided into continuous and discrete ones. The transformations which
can be obtained continuously from the identity form the proper Poincar\'e
group. A classification of irreducible representations (irreps) of the
the Poincar\'e group was given by Wigner \cite{Wig39} (see also the books
\cite{Macke68,BarRa77,Tung85,KimNo86,Ohnuk88}).
The representation theory of the proper Poincar\'e group, in fact, provides
us only by continuous transformations in the representation spaces.
At the same time, there is no a regular way to describe the discrete
transformations in such spaces on the ground of purely group-theoretical
considerations. Moreover, it turns out that there is no one-to-one
correspondence between the set ($P$,$T$) of discrete transformations in
Minkowski space and the complete set of discrete transformations in the
representation spaces. The latter set is wider than the former one (it
includes $P$,$T$,$C$,$T_w$%
 \footnote{There are three different transformations related to the
 change of the sign of time: time reflection $T$ considered in detail in
 \cite{GelMiS63}, Wigner time reversal $T_w$ \cite{Wigne32} and
 Schwinger time reversal $T_{sch}$ \cite{Schwi51,UmeKaT54}.}).

As a rule, finding discrete transformations in the representation spaces
demands an analysis of the corresponding wave equations, and has, in a sense,
heuristic character.
Besides, the possibility to have different wave equations for particles with
the same spin results in a certain "fuzziness" of the definition of
the discrete transformations in the representation spaces (see, e.g.,
\cite{LeeWi66,BenTu81}). All that stresses the lack of a regular approach
to the definition of such discrete transformations and, therefore,
creates an uncertainty in using the discrete transformations as symmetry ones.
More detail consideration have led authors of \cite{LeeWi66} to the
conclusion that "the situation is clearly an unsatisfactory one from a
fundamental point of view".

One ought to mention attempts to define the discrete transformations in the
representation spaces without appealing to any relativistic wave equations
or model assumptions. In particular, some features of the discrete
transformations were considered on the base of their commutation relations
with the generators of the Poincar\'e group \cite{Shiro58}. Using such
relations,
it is possible to define the extended Poincar\'e group, which includes the
discrete transformations as well, and then to consider irreps of the extended
group. However, here, besides of the ambiguity of such extensions
\cite{Shiro60,Wigne64,LeeWi66}, the problem of the explicit construction of
the discrete transformations in the representation spaces remains still open.

In the present work, we offer the consistent approach to the description of
the discrete transformations which is completely based on the representation
theory of the proper Poincar\'e group.
Our consideration contains two key points:

First, we study a scalar field on the proper Poincar\'e group, which carries
representations with all possible spins.
This field depends on the coordinates $x$ on Minkowski space
(which is a coset space of the Poincar\'e group with respect to the Lorentz
subgroup) and the coordinates $z$ on the Lorentz group, which correspond to
spin degrees of freedom.
Some of the discrete transformations affect only the spacetime coordinates
$x$, and some of them affect only the spin coordinates $z$.
The consideration of the scalar field on the Poincar\'e group gives a
possibility to describe "nongeometrical" transformations (i.e. ones that
leave spacetime coordinates $x$ unchanged), and, in particular, charge
conjugation, on an equal footing with the reflections in Minkowski space.
Expanding the scalar field in powers of $z$, it is easy to obtain the usual
spin-tensor fields as the corresponding coefficient functions.

Second, we identify the discrete transformations with the involutory
automorphisms of the proper Poincar\'e group.

As it is known, there are two types of automorphisms.
By definition, an inner automorphism of the group $G$ can be represented
in the form $g\to g_0gg_0^{-1}$, where $g_0\in G$. Other automorphisms are
called outer ones. It is evident that the outer automorphisms of the proper
Poincar\'e group can't be reduced to the continuous transformations of the
group (whereas for the inner automorphisms a supplementary consideration is
required); as we will see below, they correspond to the reflections of the
coordinate axes or to the dilatations.
The connection between some discrete transformations and the outer
automorphisms was also mentioned earlier
\cite{GelMiS63,Miche64,Kuo71a,Silag92,GitSh00}.
In particular, \cite{Kuo71a} contains the idea that the outer automorphisms
of internal-symmetry groups may correspond to discrete (possibly broken)
symmetries. In this context it is necessary to point out the work
\cite{GelMiS63}, where an outer automorphism of the Lorentz group was taken
as a starting point for the consideration of the space reflection
transformation.

Studying involutory (both outer and inner) automorphisms of the proper
Poincar\'e group, we describe all discrete transformations
and give explicit formulae for the action of the discrete transformations
on arbitrary spin-tensor fields without appealing to any relativistic
wave equations.

One has to point out that there is a discussion in the literature about the
sign of the mass term in the relativistic wave equations for half-integer
spins (see, e.g., \cite{Marko64,BraLj80,BarZi93,Ahluw96,Dvoeg96}).
We apply the approach under consideration to present a full solution for
such a problem.

The paper is organized as follows.

In Section 2 we show that the outer involutory automorphisms of the
Poincar\'e group are generated by the reflections in Minkowski space and,
thus, there is one-to-one correspondence between such automorphisms and
the reflections.

In Section 3 we introduce the scalar field on the Poincar\'e group and we
find transformation laws of the field under the outer and inner automorphisms.
This allows us to describe the action of all the discrete transformations
 %including space and time reflections, charge conjugation and time reversal
in terms of the field on the group.

In Section 4, decomposing the field on the group, we obtain formulae for
the action of the discrete transformations on arbitrary spin-tensor fields.

In Section 5 we derive transformation laws under the automorphisms for the
generators of the Poincar\'e group and for some other operators. On
this base, we consider properties of the discrete transformations, and, in
particular, we compare Wigner and Schwinger time reversals.

In Section 6 we extend the Poincar\'e group by the discrete transformations
and describe characteristics of irreps of the extended group.

In Sections 7,8,9 we construct explicitly massive and massless fields with
different characteristics corresponding to the discrete transformations and
consider the relation between our construction and the theory of relativistic
wave equations. Then we classify solutions of relativistic wave equations
for arbitrary spin with respect to the representations of the extended
Poincar\'e group.

%%%%%%%%%%%%%%%%%%%%%%%%%%%%%%%%%%%%%%%%%%%%%%%%%%%%%%%%%%%%%%%%%%%%%%%%%%%%%
\section{Reflections in Minkowski space and outer automorphisms
of the proper Poincar\'e group}

In this Section we consider how discrete transformations in Minkowski space
may generate outer involutory automorphisms%
 \footnote{Recall that an automorphism $A$ of a group $G$ is a mapping of the
 group onto itself which preserves the group multiplication law; for an
 involutory automorphism $A^{2}$ is the identity mapping.}
of the proper Poincar\'{e} group.

As is known, Poincar\'{e} group transformations
\be  \label{vector}
x^{\prime\mu }=\Lambda _{\;\;\nu }^\mu x^\nu +a^\mu ,
\ee
in Minkowski space ($\eta_{\mu \nu }=\mathop{{\rm diag}}(1,-1,-1,-1))$ of
coordinates $x=(x^{\mu },\;\mu =0,1,2,3)$ are defined by the pairs $%
(a,\Lambda )$, where $a=(a^{\mu })$ is arbitrary vector and the matix $%
\Lambda \in O(3,1)$. They obey the composition law
\be  \label{vector2}
(a_{2},\Lambda _{2})(a_{1},\Lambda _{1})=(a_{2}+\Lambda _{2}a_{1},\Lambda_{2}
\Lambda _{1}).
\ee
Any matrix $\Lambda$ can be presented in one of the four forms:
$\Lambda_0,\;\Lambda_s\Lambda_0,\;\Lambda_t\Lambda_0,\;
\Lambda_s\Lambda_t\Lambda_0$. Here $\Lambda_0\in SO_0(3,1)$,
where $SO_0(3,1)$ is a connected component of $O(3,1)$, and
matrices $\Lambda_s=\mathop{{\rm diag}}(1,-1,-1,-1)$,
$\Lambda_t=\diag(-1,1,1,1)$ correspond to space reflection $P$ and time
reflection $T$. Then inversion $I_x=PT=\Lambda_s\Lambda_t$. Pairs
$(a,\Lambda_{0})$ with the composition law (\ref{vector2}) form a group,
which is a semidirect product of the translation group $T(4)$ and of the
group $SO_{0}(3,1)$. We denote the latter group by $M_{0}(3,1)$.

Under space reflection the equation $x'=\Lambda_0 x+a$ takes the form
\[
\Lambda_s x'=\Lambda_s \Lambda_0 \Lambda_s^{-1}\Lambda_s x+
\Lambda_s a,\quad {\rm or}\;\;\bar x'=\overline\Lambda_0\bar{x}+\bar{a},
\]
where
\begin{equation}
\bar{x}=\Lambda _{s}x=(x^{0},-x^{k}),\quad \bar{a}=\Lambda
_{s}a=(a^{0},-a^{k}),\quad \overline{\Lambda }_{0}=\Lambda _{s}\Lambda
_{0}\Lambda _{s}^{-1}=(\Lambda _{0}^{T})^{-1}.
\end{equation}
In a similar manner, using the operations $T$ and $I_x$, we obtain finally
that $P$, $T,$ $I_x$ generate three outer involutory automorphisms of the
group $M_{0}(3,1)$:
\begin{eqnarray}
&&P:\quad (a,\Lambda _{0})\rightarrow (\bar{a},(\Lambda _{0}^{T})^{-1});
\label{autoI0x} \\
&&T:\quad (a,\Lambda _{0})\rightarrow (-\bar{a},(\Lambda
_{0}^{T})^{-1}); \\
&&I_{x}:\quad (a,\Lambda _{0})\rightarrow (-a,\Lambda _{0}).
\end{eqnarray}
Notice that $P$ and $T$ generate the same automorphism of the group $%
SO_{0}(3,1)$.

Consider now an universal covering group for $M_{0}(3,1)$. Such a group,
which we denote by $M(3,1)$, is the semidirect product of $T(4)$ and
$SL(2,C)$. As is known, there is a one-to-one correspondence between any
vectors $v$ from Minkowski space and $2\times 2$ hermitian matrices%
\footnote{%
We use two sets of $2\times 2$ matrices $\sigma_{\mu }=(\sigma_0,\sigma_k)$
and $\bar{\sigma}_{\mu }=(\sigma_{0},-\sigma_{k})$, %=\sigma^\mu$,
\be
\sigma _{0}=\left(\begin{array}{cc} 1 & 0 \\ 0 & 1 \end{array} \right),\quad
\sigma _{1}=\left(\begin{array}{cc} 0 & 1 \\ 1 & 0 \end{array} \right),\quad
\sigma _{2}=\left(\begin{array}{cc} 0 & -i \\i & 0 \end{array} \right),\quad
\sigma _{3}=\left(\begin{array}{cc} 1 & 0 \\ 0 & -1 \end{array} \right).
\ee
}
$V$ (see, e.g., \cite{BarRa77,StrWi64,BucKu95}):
\be
v^\mu \leftrightarrow V=v^{\mu }\sigma_\mu ,\quad
v^\mu = {\frac{1}{2}}{\rm Tr}(V\bar{\sigma}^{\mu }).  \label{spinor0}
\ee
Proper Poincare transformations $x'=\Lambda_{0}x+a$ can be
rewritten in new terms as
\begin{equation}
X'=UXU^{\dagger }+A,  \label{lawX}
\end{equation}
where $X=x^{\mu }\sigma _{\mu },\;$ $A=a^{\mu }\sigma _{\mu }$, and $U\in
SL(2,C)$ (two different matrices $\pm U$ correspond to any matrix $\Lambda
_{0})$. Elements of $M(3,1)$ are now given by the pairs $(A,U)$ with the
composition law
\be
(A_2,U_2)(A_1,U_1)=(U_2 A_1 U_2^{\dagger }+A_2,\; U_2 U_1).
\label{comp1}
\ee
Space reflection takes $x=(x^0,x^k)$ into $\bar{x}=(x^0,-x^k)$, or
in terms of $X=x^\mu\sigma_\mu$,
\be
P:\quad X\rightarrow \overline{X}=\bar{x}^{\mu }\sigma _{\mu }=x^{\mu }\bar{%
\sigma}_{\mu }.
\ee
Using the relation $\overline{X}=\sigma _{2}X^{T}\sigma _{2}$ and the
identity $\sigma _{2}U\sigma _{2}=(U^{T})^{-1}$, we obtain as a consequence
of (\ref{lawX})
\be
\overline{X}'=(U^{\dagger })^{-1}\overline{X}U^{-1}+\overline{A}.
\label{xbar}
\ee
Thus, $\overline{X}$ is transformed by means of the element
$(\overline{A},(U^{\dagger })^{-1})$ of $M(3,1)$. The relation
\be \label{autoPx}
P: \quad (A,U){\rightarrow }(\overline{A},(U^{\dagger })^{-1})
\ee
defines an outer involutory automorphism of the proper Poincar\'{e} group.
In a similar manner, we obtain automorphisms of the group $M(3,1)$ which are
generated by $T,I_x$,
\bea  \label{autoTx}
&&T: \quad (A,U){\rightarrow }(-\overline{A},(U^{\dagger })^{-1});
\\    \label{autoIx}
&&I_x:\quad (A,U) {\rightarrow } (-A,U).
\eea

The automorhisms corresponding to $P$ and $T$ exhaust all outer involutory
automorhisms of the Poincar\'{e} group in the following sense. Any outer
involutory automorphism can be presented as a composition of these two
automorphisms and of an inner automorphism of the group.%
 \footnote{%
 The Poincar\'{e} group $M(3,1)$ is a semidirect product of the Lorentz group
 $SL(2,C)$ and the group of four-dimensional translations $T(4)$. Any outer
 automorhism of $SL(2,C)$ is a product of involutory automorphism
 $U\rightarrow (U^{\dagger })^{-1}$ and of an inner automorphism
 \cite{GelMiS63}. Outer automorphisms of the translation group are generated
 by the dilatations $x^{\mu }\rightarrow cx^{\mu }$, $c\neq 0,1$, and are
 involutory only at $c=-1$. Outer automorphisms of $SL(2,C)$ and $T(4)$
 generate the  following outer automorphisms of the Poincar\'{e} group:
 $(A,U)\rightarrow (\overline{A},(U^{\dagger })^{-1})$,
 $(A,U)\rightarrow (cA,U)$.}
In particular, the automorphism of complex conjugation
\be  \label{autoCx}
C:\quad (A,U){\rightarrow }(\cc A,\cc U).
\ee
is the product of the outer automorphism (\ref{autoPx}) and of the inner
automorphism
\begin{equation}
(0,i\sigma ^{2})(\overline{A},(U^{\dagger })^{-1})(0,-i\sigma ^{2})=(\cc A,%
\cc U).  \label{autoCx1}
\end{equation}

As one can see from (\ref{autoPx}), (\ref{autoTx}), $P$ and $T$ generate the
same automorphisms of the Lorentz group $SL(2,C)$, namely, $U\rightarrow
(U^{\dagger })^{-1}$, whereas $PT$ generates the identity automorphism of $%
SL(2,C)$ and outer automorphism $A\rightarrow -A$ of the translation group.

Thus, we have demonstrated how descrete transformations in Minkowski space
generate outer involutory automorphisms of the proper Poincare group. In the
next Section we are going to relate the automorphisms of the proper Poincare
group with descrete transformations in the representation spaces of the group,
which are of our main interest.

%%%%%%%%%%%%%%%%%%%%%%%%%%%%%%%%%%%%%%%%%%%%%%%%%%%%%%%%%%%%%%%%%%%%%%%%%%%%%
\section{Automorphisms of proper Poincar\'e group and descrete
transformations in representation spaces}

To introduce the scalar field on the proper Poincar\'e group we describe
briefly the principal points of the corresponding technique \cite{GitSh00}.
It is well known \cite{Vilen68t,BarRa77,ZhelSc83} that any irrep of a group
$G$ is contained (up to the equivalence) in a decomposition of a generalized
regular representation. Consider the left generalized regular representation
$T_L(g)$, which is defined in the space of functions $f(h)$, $h\in G$, on the
group as
\be  \label{GRRL}
T_{L}(g)f(h)=f'(h)=f(g^{-1}h),\;g\in G.
\ee
As a consequence of the relation (\ref{GRRL}) we can write
\begin{equation}
f'(h')=f(h),\quad h'=gh.  \label{3.2}
\end{equation}
Let $G$ be the group $M(3,1)$, and we use the parametrization of its elements
by two $2\times 2$ matrices (one hermitian and another one from $SL(2,C)$),
which was described in the previous Section. At the same time, using such a
parametrization, we choose the following notations:
\begin{equation}
g\leftrightarrow (A,U),\quad  h\leftrightarrow (X,Z)\,,  \label{3.3}
\end{equation}
where $A,X$ are $2\times 2$ hermitian matrices and $U,Z\in SL(2,C).$ The map
$h\leftrightarrow (X,Z)$ creates the correspondence
\begin{eqnarray}
&&h\leftrightarrow (x,z,\uz),\quad \text{where} \quad x=(x^\mu), \;
z=(z_\alpha), \; \uz=(\uz_\alpha),
\nonumber \\ \label{3.4}
&&\mu =0,1,2,3, \; \alpha =1,2, \; z_1\uz_2-z_2\uz_1=1,
\end{eqnarray}
by virtue of the relations
\be
X=x^\mu\sigma_\mu, \quad
Z=\left(\begin{array}{cc} z_1 & \uz_{1} \\ z_2 & \uz_2 \end{array} \right)
\in SL(2,C)\,.  \label{3.5}
\ee
On the other hand, we have the correspondence
$h' \leftrightarrow (x',z',\uz')$,
\begin{eqnarray} \nonumber
&&h'=gh \leftrightarrow
(X',Z')=(A,U)(X,Z)=(UXU^{+}+A,UZ) \leftrightarrow (x',z',\uz'),
\\  \label{3.6}
&&x^{\prime\mu}\sigma_\mu=X'=UXU^{+}+A \quad \Longrightarrow \quad
 x^{\prime\mu}=(\Lambda_0)_{\;\,\nu }^\mu x^\nu +a^\mu, \quad
 \Lambda_0\leftarrow U\in SL(2,C),
\\  \label{3.7}
&&\left(\begin{array}{cc} z'_1 & \uz'_1 \\ z'_2 & \uz'_2 \end{array} \right)
 = Z'= UZ       \; \Longrightarrow \;
 z'_{\alpha}=U_{\alpha}^{\;\;\beta}z_{\beta}, \;
 \uz'_{\alpha}=U_{\alpha}^{\;\;\beta}\uz_{\beta}, \;
 U=(U_\alpha^{\;\;\beta}),\; z'_1\uz'_2-z'_2\uz'_1=1.
\end{eqnarray}
Then the relation (\ref{3.2}) takes the form
\bea
&&f'(x',z',\uz')=f(x,z,\uz),  \label{3.9}
\\ \label{3.10}
&&x^{\prime\mu}=(\Lambda_0)_{\;\,\nu}^\mu x^\nu+a^\mu, \quad
 \Lambda_0\leftarrow U\in SL(2,C),
\\ \label{lawZ}
&& z'_{\alpha}=U_{\alpha}^{\;\;\beta}z_{\beta}, \quad
 \uz'_{\alpha}=U_{\alpha}^{\;\;\beta}\uz_{\beta}, \quad
 z_1\uz_2-z_2\uz_1 = z'_1\uz'_2-z'_2\uz'_1=1.
\eea

The relations (\ref{3.9})-(\ref{lawZ}) admit a remarkable interpretation.
We may treat $x$ and $x'$ in these relations as position coordinates in
Minkowski space $M(3,1)/SL(2,C)$ (in different Lorentz refrence frames)
related by proper Poincare transformations, and the sets $z,\uz$ and
$z',\uz'$ may be treated as spin coordinates in these Lorentz frames. They
are transformed according to the formulae (\ref{lawZ}). Carrying
two-dimensional spinor representation of the Lorentz group, the variables $z$
and $\uz$ are invariant under translations as one can expect for spin degrees
of freedom. Thus, we may treat sets $x,z,\uz$ as points in a position-spin
space with the transformation law (\ref{3.10}), (\ref{lawZ}) under the change
from one Lorentz reference frame to another. In this case equations
(\ref{3.9})-(\ref{lawZ}) present the transformation law for scalar functions
on the position-spin space.

On the other hand, as we have seen, the set $(x,z,\uz)$ is in one-to-one
correspondence to the group $M(3,1)$ elements. Thus, the functions
$f(x,z,\uz)$ are still functions on this group. That is why we often call
them scalar functions on the group as well, remembering that the term
''scalar'' came from the above interpretation.

Remember now that different functions of such type correspond to different
representations of the group $M(3,1)$. Thus, the problem of classification
of all irreps of this group is reduced to the problem of a classification
of all scalar functions on position-spin space.
It is natural to restrict ourselves by the scalar functions which are
analytic both in $z,\uz$ and in $\cc z,\cc\uz$ (or, simply speaking, which
are differentiable with respect to these arguments).
Further, such functions are denoted by $f(x,z,\uz,\cc z,\cc\uz)=f(x,\bz)$,
$\bz=(z,\uz,\cc z,\cc\uz)$.
In consequence of the unimodularity of matrices $U$ there
exist invariant antisymmetric tensors
$\varepsilon^{\alpha\beta}=-\varepsilon^{\beta\alpha}$,
$\varepsilon^{\da\db}=-\varepsilon^{\db\da}$,
$\varepsilon^{12}=\varepsilon^{{\dot 1}{\dot 2}}=1$,
$\varepsilon_{12}=\varepsilon_{{\dot 1}{\dot 2}}=-1$.
Spinor indices are lowered and raised according to the rules
\be
z_\alpha=\varepsilon_{\alpha\beta}z^\beta, \quad
z^\alpha=\varepsilon^{\alpha\beta}z_\beta, \quad
\cc z_\da=\varepsilon_{\da\db}\cc z^\db, \quad
\cc z^\da=\varepsilon^{\da\db}\cc z_\db.
\ee

The continuous transformations (\ref{lawZ}) corresponding to the Lorentz
rotations are ones, which do not mix $z^\alpha$ and $\uz^\alpha$ (and their
complex conjugate $\cc z^\da$, $\cc\uz^\da$). Therefore four subspaces of
functions $f(x,z)$, $f(x,\uz)$, $f(x,\cc z)$, $f(x,\cc\uz)$ are invariant
with respect to $M(3,1)$ transformations.

In the framework of the scalar field theory on the Poincar{\'e} group
\cite{GitSh00} the standard spin description in terms of multicomponent
functions arises under the separation of space and spin variables .

Since $\bz$ is invariant under translations, any function $\phi(\bz)$
carry a representation of the Lorentz group. Let a function $f(h)=f(x,\bz)$
allows the representation
\be  \label{lor1}
f(x,\bz )= \phi^n(\bz )\psi_n(x),
\ee
where $\phi^n(\bz )$ form a basis in the representation space of the
Lorentz group. The latter means that one may decompose the functions
$\phi^n(\bz' )$ of transformed argument $\bz' =g\bz $ in terms of the
functions $\phi^n(\bz )$:
\be  \label{z_bas}
\phi^n (\bz' ) = \phi^l (\bz )L_l^{\;\;n}(U).
\ee
Thus, the action of the Poincar{\'e} group on a line
$\phi(\bz )=(\phi^n(\bz ))$ is reduced to a multiplication by matrix $L(U)$,
where $U\in SL(2,C)$: $\phi (\bz')=\phi (\bz )L(U)$.

Comparing the decompositions of the function $f'(x',\bz' )=f(x,\bz )$
over the transformed basis $\phi (\bz' )$ and over the initial basis
$\phi (\bz)$,
$$
f'(x',\bz' )=\phi (\bz' )\psi'(x')
= \phi (\bz )L(U) \psi'(x')=\phi (\bz )\psi (x),
$$
where $\psi (x)$ is a column with components $\psi_n(x)$, one may obtain
\be  \label{nfield}
\psi' (x')=L(U^{-1})\psi (x),
\ee
i.e. the transformation law of a tensor field on Minkowski space.
This law corresponds to the representation of the Poincar\'e group acting in
a linear space of tensor fields as follows
$T(g)\psi (x)=L(U^{-1})\psi (\Lambda^{-1}(x-a))$.
According to (\ref{z_bas}) and (\ref{nfield}), the functions $\phi (\bz)$
and $\psi (x)$ are transformed under contragradient representations of the
Lorentz group.

Consider now the action of automorphisms in the space of functions on the
Poincar\'e group. Automorphisms $g\to IgI^{-1}$ (both inner and outer)
generate the following transformations of the left generalized regular
representation of the Poincar\'e group:
\bea \label{autA}
&&T_L(g) \to IT_L(g)I^{-1}\equiv T_L(IgI^{-1}), \quad \\
&&f(h) \to If(h)\equiv f(IhI^{-1}), \label{autf}
\eea
where (\ref{autf}) defines the mapping of the space of functions $f(h)$ into
itself, corresponding to the automorphism (\ref{autA}). Notice that, putting
$If(h)=f(Ih)$ instead of (\ref{autf}), we come to a contradiction, since $Ih$
is not an element of the group if $I$ is an outer automorphism, and
corresponding transformation expels from the space of functions on the group.

Transformation rules of $(A,U)$ and $X$ under automorphisms corresponding
to space and time reflections are given by the formulae
(\ref{autoPx})-(\ref{autoIx}).
In order to establish the transformation rule of $Z$, it is sufficient
to note that the composition law of the group is conserved under
automorphisms, and therefore $(X,Z)$ is transformed just as $(A,U)$:
\bea  \label{autoPxz}
&&P:\quad (X,Z)\to (\overline X,(Z^\dagger )^{-1});
\\   \label{autoTxz}
&&T:\quad (X,Z)\to (-\overline X,(Z^\dagger )^{-1});
\\   \label{autoIxz}
&&I_x:\quad (X,Z)\to (-X, Z).
\eea

Thus, the automorphisms under consideration correspond to the replacement
of the arguments of scalar functions $f(h)$ on the group according to
the formulae (\ref{autoPxz})-(\ref{autoIxz}).
 %Notice that one can't insert arbitrary phase factor in right-hand side of
 %the relation $f(h)\to f(IhI^{-1})$. For example, inserting a phase factor
 %$e^{i\phi}$ and putting $f(h)=x^1$, we obtain that under space reflection
 %$x^1\to -e^{i\phi}x^1$, and therefore $e^{i\phi}=1$.

The replacement
\be \label{PZ}
Z \stackrel{P,T}{\longrightarrow} (Z^\dagger)^{-1}, \quad {\rm or} \quad
\left(\begin{array}{cc} z^1 & \uz^1 \\ z^2 & \uz^2
\end{array}\right)\;  \stackrel{P,T}{\longrightarrow} \;  \left(\begin{array}{rr}
-\cc\uz_{\dot 1} & \cc z_{\dot 1}\\ -\cc\uz_{\dot 2} & \cc z_{\dot 2}
\end{array}\right)
\ee
corresponds to space and time reflections. Transformation (\ref{PZ}) maps
functions of $z^\alpha$ into functions of $\cc\uz_{\dot\alpha}$.
Thus, the space of scalar functions on the group contains two subspaces of
functions $f(x,z, \cc\uz)$ and $f(x,\uz, \cc z)$, which are invariant ones
with respect to both transformations of the proper Poincar\'e group and the
discrete transformations under consideration (space and time reflections).
Below we will consider mainly these two subspaces, which we denote by
$V_+$ and $V_-$ respectively.

Complex conjugation
\be   \label{autoCf}
C:\quad T(g)\to C T(g)C^{-1}\equiv \cc T(g),
  \quad f(h)\to C f(h)\equiv \cc f(h),
\ee
affects both functions and complex coordinates on the Lorentz group, and
therefore it takes subspaces $V_+$ and $V_-$ into one another. The
transformation (\ref{autoCf}) of the field $f(h)$ can be identified with
charge conjugation, which interchanges particle and antiparticle fields
\cite{GitSh00}; below we consider this identification and various particular
cases in detail.

Studying involutory outer automorphisms of the Poincar\'e group and complex
conjugation in the space of functions on the group, we have obtained the
description of three independent discrete transformations (space reflection
$P$, time reflection $T$ and charge conjugation $C$). However, one can show
that there exist two supplementary discrete transformations, which are not
reduced to discussed above.

It is easy to see that we have two different transformation laws of arguments
of functions $f(h)$ under Lorentz rotations and under inner automorphisms:
\bea
&&(0,U)(X,Z)=(UXU^\dagger, UZ),
\\   \label{autIn}
&&(0,U)(X,Z)(0,U^{-1})=(UXU^\dagger, UZU^{-1}).
\eea
In both cases the coordinates $x$ are transformed the same way, and therefore
the action of inner automorphisms (\ref{autIn}) in the space of the scalar
functions $f(x)$ on Minkowski space is reduced to Lorentz rotations. But in
general case of functions $f(x,\bz)$ it is necessary to consider the action
of inner automorphisms more detail.

If inner automorphism (\ref{autIn}) corresponds to some discrete
transformation, then the conditions $U^2=e^{i\phi}$ and $\det U=1$ must be
fulfilled. Diagonal matrices with elements $e^{i\phi/2}$ and in the special
case $e^{i\phi}=-1$ also matrices of the form
\be \label{vid1}
\left(\begin{array}{cc} a & b \\ c & -a \end{array} \right), \quad a^2+bc=-1.
\ee
satisfy these conditions. Then, the square of the product of two different
matrices of the form (\ref{vid1}) also must be proportional to the identity
matrix. The latter condition reduces (up to sign) the set (\ref{vid1}) to
the set of three matrices
$$
i\sigma_1, \;i\sigma_2, \;i\sigma_3.
$$
Matrix $U=i\sigma_2$ gives an explicit realization of inner involutory
automorphism
\be   \label{auto1xz}
(X,Z)\to (\overline X^T, (Z^T)^{-1}).
\ee
(This realization we have used above, see (\ref{autoCx1}).)
A straightforward consideration of the automorphism (\ref{auto1xz}) is
inconvenient, because two coordinates $x^1,x^3$ change sign and $x^2$
remains unaltered (this correspond to the rotation by the angle $\pi$
in Minkowski space). Therefore we consider a transformation that is a
composition of the inner automorphism corresponding to the element $(0,U)$
and the Lorentz rotation corresponding to the element $(0,U^{-1})$:
\be  \label{rightU}
(X,Z) \to (X,ZU^{-1}).
\ee
For $U=i\sigma_2$ we obtain the transformation, which we denote by $I_z$,
\be \label{autoIz}
I_z:\quad (X,Z)\to(X,Z(-i\sigma_2)),\quad
\left(\begin{array}{cc} z^1 & \uz^1 \\ z^2 & \uz^2
\end{array}\right)\; {\to} \;  \left(\begin{array}{rr}
\uz^1 & -z^1\\ \uz^2 & -z^2 \end{array}\right).
\ee
This transformation maps the spaces of functions $f(x,z, \cc\uz)$ and
$f(x,\uz, \cc z)$ into one another like charge conjugation (\ref{autoCf})
but unlike the latter is reduced to replacement of arguments and does not
conjugate function.

For $U=i\sigma_3$ we obtain the transformation
\be  \label{autoIeta}
I_3:\quad
(X,Z)\to(X,Z(-i\sigma_3)),\quad
\left(\begin{array}{cc} z^1 & \uz^1 \\ z^2 & \uz^2
\end{array}\right)\; {\to} \;  \left(\begin{array}{rr}
-iz^1 & i\uz^1\\ -iz^2 & i\uz^2 \end{array}\right).
 %f(x,z,\cc\uz)\to f(x,iz,-i\cc\uz).
\ee
The transformation associated with $U=i\sigma_1$ is the product of
just considered transformations $I_z$ and $I_3$.

Thus, if in Minkowski space there exist only two independent discrete
transformations corresponding to outer automorphisms of the Poincar\'e group,
then for the scalar field on the group there exist five independent discrete
transformations corresponding to both outer and inner automorphisms,
which are not reduced to transformations of the proper Poincar\'e group.
Charge conjugation is assotiated with complex conjugation of the functions
on the group, and other four transformations are assotiated with following
replacements of the arguments of the scalar functions on the group:
\be  \label{xz}
\begin{array}{|l|rr|rrrr|}
\hline
       & x^0 & {\bf x} & z^\alpha   & \cc z_\da & \uz^\alpha &\cc\uz_\da \\
\hline
P      & x^0 &-{\bf x} &-\cc\uz_\da & \uz^\alpha & \cc z_\da &-z^\alpha \\
 %T      &-x^0 & {\bf x} &-\cc\uz_\da & \uz^\alpha & \cc z_\da &-z^\alpha \\
I_x    &-x^0 &-{\bf x} & z^\alpha   & \cc z_\da & \uz^\alpha &\cc\uz_\da \\
I_z    & x^0 & {\bf x} & \uz^\alpha & \cc\uz_\da & -z^\alpha &-\cc z_\da \\
I_3 & x^0 & {\bf x} &-iz^\alpha  & i\cc z_\da &i\uz^\alpha &-i\cc\uz_\da \\
\hline
\end{array}
\ee

%%%%%%%%%%%%%%%%%%%%%%%%%%%%%%%%%%%%%%%%%%%%%%%%%%%%%%%%%%%%%%%%%%%%%%%%%%%%%

\section{Action of the automorphisms on spin-tensor fields}

Decomposing the scalar field on the Poincar\'e group in powers of
$\bz =(z,\uz,\cc z,\cc\uz)$, it is easy to obtain transformation
laws for spin-tensor fields, which are coefficient functions and depend on
the coordinates in Minkowski space only. At the same time one must take
into account that in comparison with corresponding scalar fields on the group
it is necessary to use two sets of indices (dotted and undotted, which for
fields on the group simply duplicate the sign of complex conjugation
of the coordinates on the Lorentz group) and stipulate what kind
of object (particle or antiparticle) is described by the function (instead
of using underlined and non-underlined coordinates on the Lorentz group).

As a simple example we consider linear in $\bz$ functions, which correspond
to spin 1/2. If particle field is described by a function
$f(x,z,\cc\uz)\in V_+$,
\be
f(x,z,\cc\uz)=
\chi_\alpha(x)z^\alpha + \cc\psi^{\da}(x)\cc\uz_{\da}=Z_D\Psi(x), \quad
Z_D=(z^\alpha\; \cc\uz_{\da}), \quad
\Psi(x)={{\chi_\alpha (x)} \choose {\cc \psi^{\da}(x)} },
\label{31fdir}
\ee
then antiparticle field is described by a function $f(x,\uz,\cc z)\in V_-$,
\be
f(x,\uz,\cc z)=
\chi_\alpha(x)\uz^\alpha + \cc\psi^{\da}(x)\cc z_{\da}=\underline Z_D\Psi(x),
\quad \underline Z_D=(\uz^\alpha\; \cc z_{\da}),
\label{31fdira}
\ee
where $Z_D$ and $\underline Z_D$ (and therefore bispinor $\Psi(x)$ in both
formulae) have the same transformation law under the proper Poincar\'e group
$M(3,1)$.

According to (\ref{autoPxz}), (\ref{PZ}) we get for space reflection
$$
P:\quad Z_D\Psi(x) {\to} Z_D \Psi^{(s)}(\bar x),\quad
\Psi^{(s)}(\bar x) =-{ {\cc\psi^\da (\bar x)} \choose {\chi_\alpha (\bar x)} }
=\gamma^0\Psi(\bar x).
$$
Thus, for time reflection we have $\Psi^{(s)}(\bar x)=\gamma^0\Psi(-\bar x)$.
Charge conjugation corresponds to the complex conjugation in the space of
scalar functions, and, according to (\ref{autoCf}), we may write
\be
C:\quad Z_D\Psi(x) {\to} \cc Z_D\cc\Psi(x)=\underline Z_D \Psi^{(c)}(x),
\quad
\Psi^{(c)}(x)=-{ {      \psi_{\alpha}(x)} \choose
            { \cc\chi^{\dot\alpha} (x)} } = i\gamma^2\cc\Psi(x).
\ee
Finally, using formulae (\ref{autoIz}) and (\ref{autoIeta}), we obtain
for the transformations $I_z$ and $I_3$
\bea
&&I_z:   \quad Z_D\Psi(x) {\to} \underline Z_D \gamma^5 \Psi(x),
\\
&&I_3:\quad Z_D\Psi(x) {\to} - Z_D i \Psi(x).
\eea
Notice that both transformations $I_z$ and $C$ interchange particles and
antiparticles. The transformation $I_3$ produces only a phase factor.

In order to find transformation laws for spin-tensor fields
we need the explicit form of bases of the Lorentz group irreps. Consider the
monomial basis
$$
(z^1)^a(z^2)^b (\cc\uz_1)^c(\cc\uz_2) ^d
$$
in the space of functions $\phi(z,\cc\uz)$. The values $j_1=(a+b)/2$ and
$j_2=(c+d)/2$ are invariant under the action of generators of the Lorentz
group (\ref{SL}). Hence the space of irrep $(j_1,j_2)$ is the space of
homogeneous functions depending on two pairs of complex variables
of power $(2j_1,2j_2)$. We denote these functions as $\varphi_{j_1j_2}(z)$.

For finite-dimensional nonunitary irreps of $SL(2,C)$, $a,b,c,d$ are integer
nonnegative, therefore $j_1,j_2$ are integer or half-integer nonnegative
numbers. One can write functions $f_s(x,z)$, which are polynomials of
the power $2s=2j_1+2j_2$ in $z,\cc \uz$, in the form
\be \label{31decomp}
 f_s(x,z,\cc\uz)=\sum_{j_1+j_2=s}\sum_{m_1,m_2}
 \psi^{m_1m_2}_{j_1j_2}(x)\varphi^{m_1m_2}_{j_1j_2}(z,\cc\uz).
\ee
Here the functions
\bea \label{zbasis}
&&\varphi^{m_1m_2}_{j_1j_2}(z,\cc\uz) = N^\qq (z^1)^{j_1+m_1}(z^2)^{j_1-m_1}
 (\cc\uz_{\dot 1})^{j_2+m_2} (\cc\uz_{\dot 2})^{j_2-m_2},
\\  \label {31zbas}
&&N=(2s)![(j_1+m_1)!(j_1-m_1)!(j_2+m_2)!(j_2-m_2)!]^{-1},
\eea
form a basis of the irrep of the Lorentz group. This basis corresponds to a
chiral representation. On the other hand, one can write a decomposition of
the same functions in terms of symmetric spin-tensors
${\psi_{\alpha_1\dots\alpha_{2j_1}}}^{\db_1\dots\db_{2j_2}}(x)=
{\psi_{\alpha_{(1}\dots\alpha_{{2j_1})}}}^{\db_{(1}\dots\db_{{2j_2})}}(x)$:
\be  \label{31ten}
f_s(x,z)=\sum_{j_1+j_2=s}f_{j_1j_2}(x,z), \quad
f_{j_1j_2}(x,z)={\psi_{\alpha_1\dots\alpha_{2j_1}}}^{\db_1\dots\db_{2j_2}}(x)
z^{\alpha_1} \dots z^{\alpha_{2j_1}}\cc\uz_{\db_1} \dots \cc\uz_{\db_{2j_2}}.
\ee
Comparing these decompositions, we obtain the relation
\be
N^\qq \psi^{m_1m_2}_{j_1j_2}(x)=
  \psi_{\underbrace{\scriptstyle{1\,\dots\,1}}_{j_1+m_1}\,
        \underbrace{\scriptstyle{2\,\dots\,2}}_{j_1-m_1}}
      ^{\overbrace{\scriptstyle{\dot 1\,\dots\,\dot 1}}^{j_2+m_2}\,
        \overbrace{\scriptstyle{\dot 2\,\dots\,\dot 2}}^{j_2-m_2}}(x).
\ee

Consider now the action of the discrete transformations on the functions
$\psi(x)$. According to (\ref{autoPxz}) and (\ref{PZ}), the automorphism,
which is related to $P$, allows one to write
(see (\ref{31decomp}), (\ref{zbasis}))
\be
f(x,z,\cc\uz) \stackrel{P}{\to} f(\bar x,-\cc\uz,-z) =
\varphi(-\cc\uz,-z)\psi(\bar x) = \varphi(z,\cc\uz)\psi^{(s)}(\bar x).
\ee
It follows from (\ref{zbasis}) that
\be
\varphi^{m_1m_2}_{j_1j_2}(-\cc\uz,-z)
= (-1)^{2(j_1+j_2)}\varphi^{m_2m_1}_{j_2j_1}(z, \cc\uz),
\ee
thus we get
$$
\psi^{(s)}\vphantom{\psi}^{m_1m_2}_{j_1j_2}(\bar x)
=(-1)^{2(j_1+j_2)}\psi^{m_2m_1}_{j_2j_1}(\bar x).
$$
Finally, in terms of spin-tensor fields, we can write
\be
{\psi_{\alpha_1\dots \alpha_{2j_1}}}^{\db_1\dots \db_{2j_2}}(x)\stackrel{P}\to
(-1)^{2(j_1+j_2)}{\psi_{\beta_1\dots \beta_{2j_2}}}^{\da_1\dots \da_{2j_1}}(\bar x).
\ee
 %[for Dirac basis see \cite{GitSh00}, App.B]

Charge conjugation $C$ maps functions $f(x,z,\cc\uz)\in V_+$ into functions
$f(x,\uz,\cc z)\in V_-$:
\be
f(x,z,\cc\uz) \stackrel{C}{\to} \cc f(x,z,\cc\uz) =
\cc\varphi(z,\cc\uz) \psi^*(x) = \varphi(\uz, \cc z)\psi^{(c)}(x).
\ee
Using again (\ref{zbasis}) to write
\be
\cc\varphi^{m_1m_2}_{j_1j_2}(z,\cc\uz)
= \varphi^{m_1m_2}_{j_1j_2}(\cc z, \uz) =
(-1)^{(j_1-m_1)+(j_2+m_2)}\varphi^{m_2m_1}_{j_2j_1}(\uz, \cc z),
\ee
we obtain
$$
\psi^{(c)}\vphantom{\psi}^{m_1m_2}_{j_1j_2}(x) =
(-1)^{(j_1-m_1)+(j_2+m_2)}\cc\psi^{m_2m_1}_{j_2j_1}(x).
$$
The latter results in the following relations for spin-tensor fields
\be
{\psi_{\alpha_1\dots\alpha_{2j_1}}}^{\db_1\dots \db_{2j_2}}(x)\stackrel{C}\to
{\cc\psi^{\beta_1\dots \beta_{2j_2}}}_{\da_1\dots \da_{2j_1}}(x)=
(-1)^{(j_1-m_1)+(j_2+m_2)}
{\cc\psi_{\beta_1\dots \beta_{2j_2}}}\vphantom{\psi}^{\da_1\dots \da_{2j_1}}(x).
\ee

The action of discrete transformations on the functions
$f(x,\uz,\cc z)\in V_-$, which correspond to antiparticle fields, can be
obtained a similar way.

The obtained formulae are summarized in two tables, where we give
transformation laws both for scalar fields on the Poincar\'e group and for
spin-tensors fields in Minkowski space.

\medskip
Table 1. Discrete transformations for particle fields.
$$
\begin{array}{|l|l|l|l|}
\hline
           & f(x,z,\cc\uz)             &
{\psi_{\alpha_1\dots \alpha_{2j_1}}}^{\db_1\dots \db_{2j_2}}(x)
           &  \Psi(x)                  \\
\hline
P          & f(\bar x,-\cc\uz,-z)      &
(-1)^{2(j_1+j_2)}{\psi_{\beta_1\dots \beta_{2j_2}}}^{\da_1\dots \da_{2j_1}}(\bar x)
           &  \gamma^0\Psi(\bar x)     \\
T          & f(-\bar x,-\cc\uz,-z)     &
(-1)^{2(j_1+j_2)}{\psi_{\beta_1\dots \beta_{2j_2}}}^{\da_1\dots \da_{2j_1}}(-\bar x)
           &  \gamma^0\Psi(-\bar x)     \\
I_x=PT     & f(-x,z,\cc\uz)            &
{\psi_{\alpha_1\dots \alpha_{2j_1}}}^{\db_1\dots \db_{2j_2}}(-x)
           &  \Psi(-x)                 \\
\hline
C          & \cc f(x,z,\cc\uz)         &
{\cc\psi^{\beta_1\dots \beta_{2j_2}}}_{\da_1\dots \da_{2j_1}}(x)
           & i\gamma^2 \cc\Psi(x) \\
T_{sch}=CT & \cc f(-\bar x,-\cc\uz,-z) &
(-1)^{2(j_1+j_2)}{\cc\psi^{\alpha_1\dots \alpha_{2j_1}}}_{\db_1\dots \db_{2j_2}}(-\bar x)
           & i\gamma^0\gamma^2\cc\Psi(-\bar x)      \\
\hline
I_z        & f(x,\uz,-\cc z)           &
(-1)^{2j_2}{\psi_{\alpha_1\dots \alpha_{2j_1}}}^{\db_1\dots \db_{2j_2}}(x)
           &  \gamma^5 \Psi(x)         \\
T_w=I_zCT  & \cc f(-\bar x,\cc z,-\uz)  &
(-1)^{2j_2}{\cc\psi^{\alpha_1\dots \alpha_{2j_1}}}_{\db_1\dots \db_{2j_2}}(-\bar x)
           & -i\gamma^5\gamma^0\gamma^2\cc\Psi(-\bar x)\\
PCT_w=I_zI_x & f(-x,\uz,-\cc z)           &
(-1)^{2j_2}{\psi_{\alpha_1\dots \alpha_{2j_1}}}^{\db_1\dots \db_{2j_2}}(-x)
           &  \gamma^5 \Psi(-x)        \\
\hline
I_3     & f(x,-iz,-i\cc\uz)           &
(-i)^{2(j_1+j_2)}{\psi_{\alpha_1\dots \alpha_{2j_1}}}^{\db_1\dots \db_{2j_2}}(x)
           & -i\Psi(x)                 \\
\hline
\end{array}
$$

\medskip
Table 2. Discrete transformations for antiparticle fields.
$$
\begin{array}{|l|l|l|l|}
\hline
           & f(x,\uz,\cc z)             &
{\psi_{\alpha_1\dots \alpha_{2j_1}}}^{\db_1\dots \db_{2j_2}}(x)
           &  \Psi(x)                  \\
\hline
P          & f(\bar x,\cc z,\uz)      &
{\psi_{\beta_1\dots \beta_{2j_2}}}^{\da_1\dots \da_{2j_1}}(\bar x)
           & -\gamma^0\Psi(\bar x)     \\
T          & f(\bar x,\cc z,\uz)      &
{\psi_{\beta_1\dots \beta_{2j_2}}}^{\da_1\dots \da_{2j_1}}(-\bar x)
           & -\gamma^0\Psi(-\bar x)     \\
I_x=PT     & f(-x,\uz,\cc z)            &
{\psi_{\alpha_1\dots \alpha_{2j_1}}}^{\db_1\dots \db_{2j_2}}(-x)
           &  \Psi(-x)                 \\
\hline
C          & \cc f(x,z,\cc\uz)         &
{\cc\psi^{\beta_1\dots \beta_{2j_2}}}_{\da_1\dots \da_{2j_1}}(x)
           & i\gamma^2 \cc\Psi(x) \\
T_{sch}=CT & \cc f(-\bar x,\cc z,\uz) &
{\cc\psi^{\alpha_1\dots \alpha_{2j_1}}}_{\db_1\dots \db_{2j_2}}(-\bar x)
           & -i\gamma^0\gamma^2\cc\Psi(-\bar x)      \\
\hline
I_z        & f(x,-z,\cc\uz)           &
(-1)^{2j_1}{\psi_{\alpha_1\dots \alpha_{2j_1}}}^{\db_1\dots \db_{2j_2}}(x)
           & -\gamma^5 \Psi(x)         \\
T_w=I_zCT  & \cc f(-\bar x,\cc\uz,-z)  &
(-1)^{2j_2}{\cc\psi^{\alpha_1\dots \alpha_{2j_1}}}_{\db_1\dots \db_{2j_2}}(-\bar x)
           & -i\gamma^5\gamma^0\gamma^2\cc\Psi(-\bar x)\\
PCT_w=I_zI_x & f(-x, -z,\cc\uz)           &
(-1)^{2j_1}{\psi_{\alpha_1\dots \alpha_{2j_1}}}^{\db_1\dots \db_{2j_2}}(-x)
           & -\gamma^5 \Psi(-x)        \\
\hline
I_3     & f(x, i\uz,i\cc z)          &
i^{2(j_1+j_2)}{\psi_{\alpha_1\dots \alpha_{2j_1}}}^{\db_1\dots \db_{2j_2}}(x)
           & i\Psi(x)                   \\
\hline
\end{array}
$$

Besides of the five independent transformations $P,T,C,I_z,I_3$, we include
in these tables two operations related to the change of sign of time (Wigner
time reversal $T_w$ and Schwinger time reversal $T_{sch}$), inversion $I_x$
(which affects only spacetime coordinates $x^\mu$), and
$PCT_w$-transformation.

It is easy to see that $C^2=P^2=T^2=1$. Operators $I_z,I_3$ correspond to
products of involutory inner automorphisms and the rotation by the angle
$\pi$ (see (\ref{auto1xz})). Hence $I_z^2=I_3^2=T_w^2=R_{2\pi}$, where
$R_{2\pi}$ is the operator of rotation by $2\pi$. It changes the signs of the
spin variables,
$f(x,z,\cc\uz)\stackrel{R_{2\pi}}{\longrightarrow}f(x,-z,-\cc\uz)$. The
latter corresponds to the multiplication by the phase factor
$(-1)^{2(j_1+j_2)}$ only.

In the general case the transformation laws for particle and antiparticle
spin-tensor fields are distinguished by signs (for space reflection this
fact is pointed out, in particular, in \cite{Sachs87}).
This signs play an important role, because their change leads to
non-commutativity of discrete transformations.

There are two different transformations $C$ and $I_z$ which interchange
particle and antiparticle fields. The operator $I_z$ is a spin part
of $PCT_w$-transformation. Indeed, the relation $PCT_w=I_xI_z$ means that
$PCT_w$-transformation is factorized in inversion $I_x$, affecting only
spacetime coordinates $x^\mu$ and in $I_z$-transformation, affecting only
spin coordinates $\bz$.

Consider now scalar fields which are eigenfunctions for $C$; they describe
neutral particles. Such fields obey the condition
$Cf(h)=\cc f(h)=e^{i\phi} f(h)$. Multiplying them by the phase factor
$e^{i\phi/2}$ we transform them to real fields obeying the condition
$Cf(h)=f(h)$. The charge conjugation $C$ maps $z,\uz$ into the complex
conjugate pair. Thus, there are two invariant (with respect to $C$) subspaces
of the scalar functions, namely, spaces of real functions $f(x,z,\cc z)$
and $f(x,\uz,\cc\uz)$, which we denote by $V_z$ and $V_\uz$ respectively.
They are mapped into one another under space reflection,
$V_z\stackrel{P}\leftrightarrow V_\uz$. Linear in $z,\cc z$ eigenfunctions of
$C$ (with the eigenvalue 1) have the form
\bea \nonumber
&&f(x,z,\cc z)=
 \psi_\alpha(x)z^\alpha - \cc\psi^{\da}(x)\cc z_{\da}=Z_M\Psi_M(x),
\\
&&Z_M=(z^\alpha\; \cc z_{\da}), \quad
 \Psi_M(x)={{\psi_\alpha (x)} \choose {-\cc\psi^{\da}(x)} },
\label{31fmaj}
\eea
where $\Psi_M(x)$ is a Majorana spinor, $\cc\Psi_M(x)=i\gamma^2\Psi_M(x)$.
The space reflection maps functions from $V_z$ into functions from $V_\uz$,
\bea  \nonumber
P:\quad && Z_M\Psi_M(x) {\to} \underline Z_M \Psi^{(s)}_M(\bar x),\quad
\underline Z_M=(\uz^\alpha\; \cc\uz_\da), \quad \\
&&\Psi^{(s)}_M(\bar x) =-{ {\cc\psi^\da (\bar x)} \choose {\psi_\alpha (\bar x)} }
=\gamma^0\gamma^5\Psi_M(\bar x)=i\gamma^0\gamma^5\gamma^2\cc\Psi_M(\bar x).
\eea
Therefore, the spaces $V_z$ and $V_\uz$ (in contrast to the spaces $V_+$ and
$V_-$) do not contain eigenfunctions of $P$ (i.e. states with definite
parity). According to (\ref{autoIz}) and (\ref{autoIeta}) we obtain
\bea
&&I_z: \quad Z_M\Psi(x) {\to} \underline Z_M \Psi(x),
\\
&&I_3: \quad Z_M\Psi(x) {\to} - Z_M i \gamma^5\Psi(x).
\eea
Thus, there are four nontrivial independent discrete transformations for the
fields under consideration. These transformations for bispinors $\Psi(x)$ and
$\Psi_M(x)$ are performed by matrices from the same set.
However, one and the same discrete symmetry operation induces different
operations with bispinors $\Psi(x)$ and $\Psi_M(x)$.

The $PCT_w$-transformation maps the spaces of functions $f(x,z,\uz)$ and
$f(x,\cc z,\cc\uz)$ into themselves. Such functions can be used, in particular,
for describing "physical" Majorana particle defined as
$PCT_w$-self-conjugate particle with spin 1/2 \cite{KayGo83}.

%%%%%%%%%%%%%%%%%%%%%%%%%%%%%%%%%%%%%%%%%%%%%%%%%%%%%%%%%%%%%%%%%%%%%%%%%%%%%

\section{Transformational laws of operators}

Consider the action of the involutory automorphisms, which correspond to the
discrete transformations, in terms of Lee algebra of the Poincar\'e group.
Generators of the Poincar\'e group in the left generalized regular
representation have the form
\be  \label{Lgen}
\hat{p}_{\mu }=-i\partial /\partial x^{\mu }, \quad
 \hat{J}_{\mu\nu }= \hat{L}_{\mu\nu }+ \hat{S}_{\mu \nu},
\ee
where $\hat{L}_{\mu\nu }=i(x_\mu\partial _\nu -x_\nu\partial _\mu )$
are orbital momentum operators and $\hat{S}_{\mu\nu }$ are spin
operators depending on $\bz$ and $\partial/\partial \bz$. Explicit form
of the spin operators is given in the Appendix.
The generators (\ref{Lgen}) obey the commutation relations
\bea
&&[\hat{p}_{\mu },\hat{p}_{\nu }]=0,\quad  [\hat{J}_{\mu\nu},\hat{p}_{\rho}]
 =i(\eta_{\nu \rho}\hat{p}_{\mu} -\eta_{\mu \rho}\hat{p}_{\nu}), \quad
\nonumber \\
&&[\hat J_{\mu \nu },\hat J_{\rho \sigma }] =i\eta _{\nu \rho }\hat J_{\mu\sigma }
 -i\eta _{\mu \rho }\hat J_{\nu \sigma }-i\eta _{\nu \sigma }\hat J_{\mu \rho}
 +i\eta _{\mu \sigma }\hat J_{\nu \rho }\;. \label{komm0}
\eea

Fields on the Poincar\'e group depend on 10 independent variables. For the
classification of these fields one can use a complete set of commuting
operators on the group, which along with the left generators (\ref{Lgen})
includes generators in the right generalized regular representation
\be \label{GRRR}
T_{R}(g)f(x,z) = f(xg,\; zg),\quad xg \leftrightarrow X + ZAZ^{\dagger},
\quad  zg \leftrightarrow ZU.
\ee
As a consequence of the formula (\ref{GRRR}) one can obtain
\be  \label{Rgen}
\hat{p}_{\mu }^R \bar\sigma^\mu
= -Z^{-1}\hat{p}_{\nu }\bar\sigma^\nu (Z^\dagger)^{-1}, \quad
\hat{J}_{\mu\nu }^R= \hat{S}_{\mu \nu}^R.
\ee
Operators $\hat{S}_{\mu\nu}$ and $\hat{S}^R_{\mu\nu }$ from (\ref{Lgen}) and
(\ref{Rgen}) are the left and right generators of $SL(2,C)$ and do not depend
on $x$. All the right generators (\ref{Rgen}) commute with all the left
generators (\ref{Lgen}) and obey the same commutation relations (\ref{komm0}).
Below for spin projection operators we use three-dimensional vector notation
$\hat S_k =\frac 12\epsilon_{ijk} \hat S^{ij}$, $\hat B_k =\hat S_{0k}$.
The explicit form of the spin operators is given by formulae
(\ref{SL})-(\ref{31S}).

According to theory of harmonic analysis on Lie groups
\cite{ZhelSc83,BarRa77} there exists a complete set of commuting operators,
which includes Casimir operators, a set of the left generators and a set
of right generators (both sets contain the same number of the generators).
The total number of commuting operators is equal to the number of
parameters of the group. In a decomposition of the left GRR the
nonequivalent representations are distinguished by eigenvalues of the
Casimir operators, equivalent representations are distinguished by
eigenvalues of the right generators, and the states inside the irrep are
distinguished by eigenvalues of the left generators.

The physical meaning of right generators usually is not so transparent as
of left ones. However, the right generators of $SO(3)$ in the
nonrelativistic rotator theory are interpreted as angular momentum operators
in a rotating body-fixed reference frame \cite{Wigne59,LanLi3,BieLo81}.
Since the right transformations commute with the left ones, they define
quantum numbers, which are independent of the choice of the laboratory
reference frame.

The right generators $\hat S^3_R$ and $\hat B^3_R$ of the Poincar\'e group
can be used to distinguish functions from the subspaces $V_+,V_-$ and
$V_z,V_\uz$. Polynomials of power $2s$ belonging to $V_+$ and $V_-$ are
eigenfunctions of the operator $\hat S^3_R$ with eigenvalues $\mp s$
respectively. Polynomials of power $2s$ belonging to $V_z$ and $V_\uz$
are eigenfunctions of the operator $i\hat B^3_R$ with eigenvalues $\mp s$
respectively.

The explicit form of the generators (see (\ref{Lgen}),(\ref{Rgen}) and
(\ref{SL}),(\ref{SR})) allows us easily to establish their transformation
properties under involutory automorphisms, and thus under discrete
transformations. The transformations $P,T$ correspond to the outer
automorphisms of the algebra. Therefore the left and the right generators
have identical transformation rules under $P$ and $T$; in particular,
$$
\hat p_\mu\to \mp(-1)^{\delta_{0\mu}}\hat p_\mu,
\;\hat {\bf S}\to\hat {\bf S},\;\hat {\bf B}\to-\hat {\bf B},
$$
where the upper sign corresponds to $P$. Obviously, spatial and boost
components of total and orbital angular momenta have the same transformation
rules that $\hat {\bf S}$ and $\hat {\bf B}$.

Complex conjugation $C$ leads to the change of sign of all generators, as
it follows from their explicit form. Signs in the commutation relations are
also changed, and for their restoring it is necessary to replace $i$ by $-i$
in (\ref{komm0}).

The transformations $I_z,I_3$, connected with inner automorphisms, according
to (\ref{rightU}) are defined as right finite transformations of the
Poincar\'e group. They do not affect the left generators because the right
transformations commute with the left ones.
Thus the transformations $I_z,I_3$ induce automorphisms of the algebra of
the right generators: $I_z$ changes signs of the first and the third
components of $\hat{\bf S}_R$ and $\hat{\bf B}_R$, and $I_3$ changes signs
of the first and the second components of $\hat{\bf S}_R$ and $\hat{\bf B}_R$.

An intrinsic parity of a massive particle is defined as the an eigenvalue of
the operator $P$ in the rest frame, $P f(h)=\eta f(h)$, $\eta=\pm 1$.
Since the operator $P$ commutes with $T,C,I_z$, the intrinsic parity is not
changed under corresponding discrete transformations.

Transformation properties of physical quantities under the discrete
transformations are represented in Table 3. To compose this table we have
used Tables 1,2 and the explicit form of corresponding operators given in the
Appendix. The intrinsic parity $\eta$ and the sign of $p_0$ label irreps of
the improper Poincar\'e group, which include space reflection; farther
the table includes the left generators $\hat p_\mu$, the spin parts
$\hat {\bf S}$, $\hat {\bf B}$ of the left Lorentz generators, and two
right Lorentz generators.

We also include a current four-vector $j_\mu$ for the first order equation
(\ref{fseq0}) (the Dirac and Duffin-Kemmer equations are the particular cases
of this equation for $s=1/2$ and $s=1$ respectively). In the space of
functions on the group this current is described by the operators
$\hat\Gamma_\mu$, see (\ref{31Gmu}). The particle and antiparticle fields are
distinguished by the sign of charge, i.e. by the sign of a component $j_0$ of
the current vector. One can see from the table, that for scalar fields on the
group the sign of the right generator $S^3_R$ can be used to distinguish
particles and antiparticles, because this sign and the sign of $j_0$ have
identical transformation rules under the discrete transformations.
As we will show, the sign of the mass term in the equation (\ref{fseq0})
is changed with the sign of the product $p_0 S_R^3$ under discrete
transformations, see the next to last column of Table 3.

\medskip
Table 3. The action of the discrete transformations on the signs of
physical quantities.

$$
\begin{array}{|l|c|cl|rc|cc|cc|c|c|}
\hline
 &\; \eta \;&\; p_0 \;& {\bf p} \;&\;{\bf S} &{\bf B}\; &\; S_R^3 & B_R^3\;
 &\; {j_0}^{\,1)} &{\bf j}^{\,1)}\; &\; p_0 S_R^3 \;& \text{L-R} \\
\hline
P             & + & + &  - & + & -  & + & - & + & -  &+ &- \\
T             & + & - &  + & + & -  & + & - & + & -  &- &- \\
I_x=PT        & + & - &  - & + & +  & + & + & + & +  &- &+ \\
\hline
C             & + & - &  - & - & -  & - & - & - & -  &+ &- \\
CP            & + & - &  + & - & +  & - & + & - & +  &+ &+ \\
T_{sch}=CT    & + & + &  - & - & +  & - & + & - & +  &- &+ \\
PCT=I_xC      & + & + &  + & - & -  & - & - & - & -  &- &- \\
\hline
I_z           & + & + &  + & + & +  & - & - & - & -  &- &+ \\
T_w=I_zCT     & + & + &  - & - & +  & + & - & + & -  &+ &+ \\
PT_w=I_zI_xC  & + & + &  + & - & -  & + & + & + & +  &+ &- \\
PCT_w=I_zI_x  & + & - &  - & + & +  & - & - & - & -  &+ &+ \\
\hline
\multicolumn{11}{l}
{\hbox {\small ${}^{1)}$ for the states described by the first order equation
(\ref{fseq0})}} \\
\end{array}
$$

The last column of the table ("L-R") describe the passage between two
types of spinors (left $\cc z_\da$, $\cc\uz_\da$ and right $z^\alpha$,
$\uz^\alpha$) labelled by dotted and undotted indices. If the transformation
interchanges dotted and undotted spinors, then this column contains sign
"$-$", and contains sign "$+$" in the opposite case. If we define a chirality
as the difference between the number of dotted and undotted indices, then the
last column of the table corresponds to the sign of the chirality. In the
space of functions on the group the chirality is described by the operator
$\hat\Gamma^5$, see (\ref{31G5}).

The time reflection transformation $T$ (in context of the Lorentz group it
was considered in detail in \cite{GelMiS63}) maps positive energy states into
negative energy ones. On the other hand, time reversal is defined usually by
the relation $x\to-\bar x$ with supplementary condition of conservation of
the energy sign. Obviously, the product of time reflection and charge
conjugation $CT$, which we denote by $T_{sch}$, obeys this condition.
The transformation $T_{sch}$ was introduced by Schwinger \cite{Schwi51},
see also \cite{UmeKaT54}. This transformation interchanges particle and
antiparticle fields with opposite sign of the component $j_0$ of the current
vector.

The time reversal transformation $T_w$ was considered for the first time
by Wigner \cite{Wigne32}. Wigner time reversal conserves the sign of $j_0$.
Connecting different states of the same particle, this transformation
is an analog of time reversal in nonrelativistic quantum mechanics.
Changing signs of the vectors ${\bf p},{\bf S},{\bf j}$, Wigner time
reversal corresponds to the reversal of the direction of motion.
(Notice that sometimes the term "time reversal" is used instead of
"time reflection" also for transformations changing the sign of energy,
which can lead to misunderstanding.)

The transformation $I_z$ is the right finite transformation of the proper
Poincar\'e group, see (\ref{rightU}). This transformation leaves signs of
the left generators unaltered (because the right transformations commute
with the left ones) but changes signs of the current vector and of some right
generators. Hence, the left generators have identical transformation rules
under $T_{sch}$ and $T_w=I_zT_{sch}$.
The transformation $I_3$, as was mentioned above, change the sign of
the first and second components of vectors ${\bf S}_R$ and ${\bf B}_R$ only
and conserve signs of all quantities contained in Table 3.

%%%%%%%%%%%%%%%%%%%%%%%%%%%%%%%%%%%%%%%%%%%%%%%%%%%%%%%%%%%%%%%%%%%%%%%%%%%%%%
\section{Representations of the extended Poincar\'e group}

Consider the problem of extending the Poincar\'e group by means of
the discrete transformations.

Different fields on the Poincar\'e group with identical transformation
rule under left transformations (i.e. under rotations and translations of
the reference frame) carry equivalent subrepresentations of the left
generalized regular representation (\ref{GRRL}), even if they have different
transformation rules under right transformations (\ref{GRRR}).
The discrete transformations (automorphisms) also act in the space of
functions on the group, and functions carrying equivalent representations of
the proper Poincar\'e group can be transformed differently under discrete
automorphisms. Therefore, these functions carry non-equivalent representations
of the Poincar\'e group extended by the discrete transformations.

The operator $I$, $I^2=1$, corresponding to a discrete transformation and
the identity operator form finite group $Z_2$ consisting of two elements.
The operator $I$ allows us to distinguish two types of states, one with
definite "charge" and another with definite "charge parity".
Two states with opposite "charges", which we denote by $\psi_+$ and $\psi_-$,
are interchanged under the reflection $I$:
$\psi_+\stackrel{I}{\leftrightarrow}\psi_-$.
The states $\psi_+\pm\psi_-$ with definite "charge parity" are eigenfunctions
of operator $I$ with the eigenvalues $\pm 1$ and form the bases of
one-dimensional irreps of the group $Z_2$. The operators $(1\pm I)/2$
obey the condition $((1\pm I)/2)^2=(1\pm I)/2$ and thus they are projection
operators on the states with definite "charge parity".

The operators of the discrete transformations (automorphisms) commute with each
other and commute (sign "$+$" in Table 3) or anticommute (sign "$-$"
in Table 3) with the generators of the Poincar\'e group. The latter means
that the discrete transformation interchanges eigenfunctions of the
generator with opposite eigenvalues (opposite "charges").

Here for clearness we adduce the table with parameters labelling the
Poincar\'e group irreps which correspond to finite-component (with respect
to spin) massive and massless fields.

\medskip
Table 4. Parameters labelling irreps of the proper and of improper
Poincar\'e groups.

$$
\begin{array}{|l|c|c|}
\hline
 &\text{proper Poincar\'e group}&
 \begin{array}{l} \text{improper Poincar\'e group} \\
 \text{with space reflection} \end{array}\\
\hline
\text{massive case} & m,\sign p_0, s & m,\sign p_0, s,\eta \\
\hline
\text{massless case} & \sign p_0,\lambda & \sign p_0,|\lambda|,\eta^{\,1)} \\
\hline
\multicolumn{3}{l}
{\hbox {\small ${}^{1)}$ for $\lambda\ne 0$ massless irreps with $\eta=\pm 1$
are equivalent \cite{ShaLe74,Tung85}}}
\end{array}
$$

\noindent
Here the mass $m>0$, the spin $s=0,1/2,1,\dots$, the intrinsic parity
$\eta=\pm 1$, and the helicity $\lambda=0,\pm 1/2,\pm 1,\dots$.
Mass $m$ and sign of $p_0$ label the orbit (the upper or lower sheet of
hyperboloid or cone), $s$ and $\lambda$ label irreps of the little groups
$SO(3)$ and $SO(2)$, $s,\eta$ and $|\lambda|,\eta$ label irreps of the little
groups $O(3)$ and $O(2)$ respectively (see \cite{Macke68,Tung85} for details).
The mass and the spin can by also defined as eigenvalues of the Casimir
operators:
\be \label{pw}
\hat p^2 f(x,\bz) = m^2 f(x,\bz), \quad
\hat W^2 f(x,\bz) = -m^2s(s+1)f(x,\bz),
\ee
where $\hat W^\mu$ is Lubanski-Pauli four-vector,
$\bz =(z,\uz,\cc z,\cc\uz)$ are coordinates on the Lorentz group.

In order to find parameters labelling irreps of the extended Poincar\'e
group, we consider four independent discrete transformations: $P,I_x,I_z,C$.

1. Irreps of the improper Poincar\'e group including space reflection $P$ can
be classified (as it was mentioned above) with the help of the little group
method. Space reflection allows us to distinguish two types of the states:
ones with definite intrinsic parity $\eta$ and ones with definite (left or
right) charge, or chiral states. In the space of functions on the group the
chirality operator $\hat \Gamma^5$ is given by the formula (\ref{31G5}).
Fields with zero chirality can be considered as pure neutral ones with
respect to space reflection.

2. Inversion $I_x$ affects only spacetime coordinates $x$. It couples two
irreps of the proper (or improper) Poincar\'e group characterized by
$\sign p_0=\pm 1$ into one representation of extended group.
Eigenstates of $I_x$ are the states with definite "energy parity".
However, since the sign of energy $p_0$ is already used to label irreps of
the proper group, this extension does not lead to appearing a supplementary
characteristic.

3. As mentioned earlier, the operator $I_z$ is the spin part of
$PCT_w$-transformation, $PCT_w=I_xI_z$, and affects only spin coordinates
$\bz$. Operator $I_z$ commutes with all the left generators and with space
reflection $P$, and thus can't change the parameters labelling irreps of the
proper or improper Poincar\'e group. $I_z$ interchanges the states with
opposite eigenvalues of $\hat S^3_R$; a charge parity $\eta_c=\pm 1$ arises
as its eigenvalue. Corresponding extension of the group, conserving all
characteristics from Table 4, gives, in addition, the charge parity
$\eta_c$ as a characteristic of irreps. Below, taking into account the simple
relation of $I_z$ to $PCT_w$-transformation, we will call corresponding
quantities by $PCT_w$-charge and $PCT_w$-parity.

4. Charge conjugation $C$ changes signs of all the generators.
In fact, this means that any extension connected with different sets
of the discrete transformations including $C$ must be considered separately.
Here we note that $C$ does not change $\eta$ and $\eta_c$ and like $I_z$
changes sign of the charge $S^3_R$.
However, if the Poincar\'e group is already extended by $P,T,I_z$, then
instead of $C$ one can consider Wigner time reversal $T_w=I_zCT$ as fourth
independent transformation. The latter corresponds to the reversal of the
direction of motion and does not change $P$-charge (chirality) and
$I_z (CPT_w)$-charge.

As a result, considering four independent discrete transformations, we have
established that irreps of the extended Poincar\'e group have two
supplementary characteristics with respect to irreps of the proper group: the
intrinsic parity $\eta$ and $PCT_w$-parity $\eta_c$. These characteristics are
associated with $P$-charge (chirality) and $PCT_w$-charge (in particular it
distinguishes particles and antiparticles), which in the space of functions
on the group are defined as eigenvalues of the operators $\hat \Gamma^5$ and
$\hat S^3_R$. It is necessary to note that for the particles with
half-integer spins these charges also are half-integer.
This means that for half-integer spins there are no pure neutral
(with respect to the discrete transformations under consideration) particles,
which have zero chirality or zero $PCT_w$-charge. On the other hand,
both for integer and half-integer spins it is possible to construct
states with definite intrinsic parity (e.g., the Dirac field)
or with definite $PCT_w$-parity (e.g., "physical" Majorana field),
which are mapped into themselves under $P$ and $PCT_w$ respectively.

%%%%%%%%%%%%%%%%%%%%%%%%%%%%%%%%%%%%%%%%%%%%%%%%%%%%%%%%%%%%%%%%%%%%%%%%%%%%%%
\section{Discrete symmetries of the relativistic wave equations. Massive case}

Here we explicitly construct the massive fields on the Poincar\'e group
and analyze their characteristics associated with the discrete
transformations. On this base we give a compact group-theoretical
derivation of basic relativistic wave equations and consider their discrete
symmetries. In particular, this allow us to solve an old problem concerning
two possible signs of a mass term in the Dirac equation.
We also give a classification of the solutions of various types of higher
spin relativistic wave equations with respect to characteristics of the
extended Poincar\'e group irreps; this classification is turned out to be
nontrivial, especially for the case of first order equations.

Consider eigenfunctions of operators $\hat p_\mu$ (plane waves).
For $m\ne 0$ there exists the rest frame, where the dependence on $x$
is reduced to the factor $e^{\pm imx^0}$.
Linear functions of coordinates $\bz$ on the Lorentz group correspond to
spin 1/2. For fixed mass $m$ there are 16 linearly independent functions:
\be \label{LRm}
\ba{lcc}
        &   V_+ &  V_- \\
L:\quad & e^{\pm imx^0}z^\alpha \quad   & e^{\pm imx^0}\uz^\alpha  \\
R:\quad & e^{\pm imx^0}\cc\uz_\da \quad & e^{\pm imx^0}\cc z_\da   \\
\ea
\ee
They can by classified (labelled) by means of left generators of the
Poincar\'e group and the operators of the discrete transformations $P,C$.
The eigenvalues of the left generators $\hat J^3$ (in the rest frame
$\hat J^3=\hat S^3$) and $\hat p^0$ give the spin projection
(for $\alpha,\da=1$ and $\alpha,\da=2$ we have $s^3=\pm 1/2$ respectively)
and the sign of $p_0$. This sign, along with the mass $m$ and the spin $s$,
characterizes nonequivalent irreps of the proper Poincar\'e group.
The operator $I_x$ interchanges the states with opposite signs of $p_0$,
the operator $P$ interchanges $L$- and $R$-states (states with opposite
chiralities), and the operators $C$ and $I_z$ interchange the particle and
antiparticle states, belonging to the spaces $V_+$ and $V_-$ respectively.
Unlike the charge conjugation operator $C$ the operator $I_z$ conserves signs
of energy and chirality.

However, the states (\ref{LRm}) with definite chirality are transformed under
reducible representation of the improper Poincar\'e group, whose irreps are
characterized by intrinsic parity $\eta$. In the rest frame the states with
definite $\eta$ are eigenfunctions of the operator $P$,
$P f(x,\bz)=\eta f(x,\bz)$:
\be \label{etam}
\begin{array}{lcc}
        &   V_+ &  V_- \\
\eta=-1:\quad  & e^{\pm imx^0}(z^\alpha + \cc \uz_\da) \quad &
e^{\pm imx^0}(\uz^\alpha - \cc z_\da) \\
\eta=1:\quad & e^{\pm imx^0}(z^\alpha - \cc \uz_\da) \quad &
e^{\pm imx^0}(\uz^\alpha + \cc z_\da) \\
\end{array}
\ee
As in the case of the states (\ref{LRm}), the operators $C$ and $I_z$
interchange functions from the spaces $V_+$ and $V_-$.
On the other hand, the states with different intrinsic parity $\eta=\pm1$
(unlike the states with different chirality) are not interchanged by
the operators of the discrete transformations.

Both the states (\ref{LRm}) and (\ref{etam}) are eigenstates of
the Casimir operators $\hat p^2$ and $\hat W^2$ of the Poincar\'e group
with the eigenvalues $m^2$ and $-(3/4)m^2$. But only the states (\ref{etam})
are transformed under irrep of the improper Poincar\'e group. Besides, it is
easy to check that the states (\ref{etam}) (unlike the states (\ref{LRm}))
are the solutions of the equations
\be  \label{pmm}
(\hat p_\mu\hat\Gamma^\mu \pm ms)f(x,z,\cc\uz) = 0, \quad
(\hat p_\mu\hat\Gamma^\mu \mp ms)f(x,\uz,\cc z) = 0,
\ee
where $s=1/2$, upper sign corresponds to $\eta\sign p_0=1$, and lower sign
corresponds to $\eta\sign p_0=-1$. The operator $\hat p_\mu\hat\Gamma^\mu$
(explicit form of $\hat\Gamma^\mu$ is given by (\ref{31Gmu})) is not
affected by space inversion and charge conjugation. The spaces $V_+$ and
$V_-$ also are invariant under space reflection, but they are interchanged
under charge conjugation.

Considering the action of the discrete transformations on the component $j_0$
of current four-vector of free equation, we have established above that if
particle is described by the function from the space $V_+$, then antiparticle
is described by the function from the space $V_-$. This can be also shown on
the base of the equations in an external field. Acting by the operator of
charge conjugation $C$ (which in the space of functions on the group acts as
the operator of complex conjugation) on the equation
\be  \label{A-eq}
((\hat p_\mu-eA_\mu(x))\hat\Gamma^\mu \pm ms)f(x,z,\cc\uz) = 0
\ee
for functions $f(x,z,\cc\uz)\in V_+$, we obtain that the functions
$\cc f(x,z,\cc\uz)\in V_-$ obey the equation with opposite charge:
\be  \label{A+eq}
((\hat p_\mu+eA_\mu(x))\hat\Gamma^\mu \pm ms)\cc f(x,z,\cc\uz) = 0.
\ee

Substituting the functions $f(x,z,\cc\uz)=Z_D\Psi(x)$ and
$\cc f(x,\uz,\cc z)=\underline Z_D\Psi^{(c)}(x)$ into equations (\ref{A-eq})
and (\ref{A+eq}) (see also (\ref{31fdir}), (\ref{31fdira})), we obtain two
Dirac equations for charge-conjugate bispinors $\Psi(x)$ and $\Psi^{(c)}(x)$:
$$
((\hat p_\mu-eA_\mu(x))\gamma^\mu \pm m)\Psi(x) = 0, \quad
((\hat p_\mu+eA_\mu(x))\gamma^\mu \pm m)\Psi^{(c)}(x) = 0.
$$

Thus we have to use the different scalar functions on the group to describe
particles and antiparticles and hence two Dirac equations for both signs of
charge respectively. That matches completely with the results of the article
\cite{GavGi00}. It was shown there that in the course of a consistent
quantization of a classical model of spinning particle namely such
(charge symmetric) quantum mechanics appears. At the same time it
is completely equivalent to the one-particle sector of the corresponding
quantum field theory.

In Section 8 we continue to consider spin-1/2 case and give an exact
group-theoretical formulation of conditions which lead to the Dirac equation.

In general case for the classification of functions corresponding to higher
spins it is necessary to use a complete  set of ten commuting operators
(including also right generators) on the group, for example
\be  \label{31set}
\hat p_\mu,\; \hat W^2,\; \hat{\bf p}\hat{\bf S}, \;
\hat {\bf S}^2 -\hat {\bf B}^2, \; \hat{\bf S}\hat{\bf B}, \;
\; \hat S^3_R,\; \hat B^3_R.
\ee
In the rest frame $\hat{\bf p}\hat{\bf S}=0$, and the complete  set can be
obtained from (\ref{31set}) by changing $\hat{\bf p}\hat{\bf S}$ to
$\hat S^3$. Functions from the spaces $V_+$ and $V_-$ depend on eight
real parameters, and therefore one can consider only eight operators;
our choice is
\be  \label{31set8}
\hat p_\mu,\; \hat W^2,\; \hat{\bf p}\hat{\bf S}\;
(\hat S^3\text{ in the rest frame}),\; \hat p_\mu\hat\Gamma^\mu,\; \hat S_3^R.
\ee
The problem of constructing the complete sets of the commuting operators on the
Poincar\'e group was discussed in \cite{Hai69,BarRa77,GitSh00}.

Consider eigenfunctions of the operators (\ref{31set8}).
For functions from the spaces $V_+$ and $V_-$ one can show that if the
eigenvalue of $\hat p_\mu\hat\Gamma^\mu$ is equal to $\pm ms$, where $2s$ is
the power of polynomial (the eigenvalue of $\hat S^3_R$), then the eigenvalue
of the operator $\hat W^2$ is also fixed and corresponds to spin $s$
\cite{GitSh00}. Thus, the system
\be  \label{sys1}
\hat p^2 f(x,\bz)=m^2 f(x,\bz),\quad
\hat p_\mu\hat\Gamma^\mu f(x,\bz)=\pm ms f(x,\bz),\quad
\hat S^3_R f(x,\bz)=\pm s f(x,\bz),
\ee
picks out the states with definite mass and spin.

Depending on the choice of the functional space and of the sign of the mass
term, the second equation of the system (\ref{sys1}) can be written in one
of four forms:
\bea
&& \qquad\qquad\qquad  V_+:  \qquad\qquad\qquad\qquad\quad  V_-: \nonumber \\
&&(\hat p_\mu\hat\Gamma^\mu + ms)f(x,z,\cc\uz) = 0 \qquad
(\hat p_\mu\hat\Gamma^\mu - ms)f(x,\uz,\cc z) = 0          \label{eqpm}\\
&&(\hat p_\mu\hat\Gamma^\mu - ms)f(x,z,\cc\uz) = 0 \qquad
(\hat p_\mu\hat\Gamma^\mu + ms)f(x,\uz,\cc z) = 0          \label{eqpm1}
\eea
In the rest frame for definite $m$ and $s$ the solutions of equations
(\ref{eqpm}) and (\ref{eqpm1}) are given by (\ref{m1}) and (\ref{m2})
respectively:
\bea  \nonumber
        &   V_+: & \qquad\qquad\qquad\quad  V_-: \\
\eta\sign p_0=-1: \;\,  &  e^{\pm imx^0}
(z^1\pm \cc\uz_{\dot 1})^{s+s^3}(z^2\pm \cc\uz_{\dot 2})^{s-s^3} & \quad
e^{\mp imx^0}(\uz^1\pm\cc z_{\dot 1})^{s+s^3}(\uz^2\pm\cc z_{\dot 2})^{s-s^3}
\label{m1}  \\  \label{m2}
\eta\sign p_0=1: \quad  &  e^{\mp imx^0}
(z^1\pm \cc\uz_{\dot 1})^{s+s^3}(z^2\pm \cc\uz_{\dot 2})^{s-s^3} & \quad
e^{\pm imx^0}(\uz^1\pm\cc z_{\dot 1})^{s+s^3}(\uz^2\pm\cc z_{\dot 2})^{s-s^3}
\eea
Here the sign of $\eta p_0$ is specified for half-integer spins; for integer
spins all solutions are characterized by $\eta=1$.
These solutions are eigenfunctions of the Casimir operators $\hat p^2$ and
$\hat W^2$ with the eigenvalues $m^2$ and $-s(s+1)m^2$ and of spin projection
operator $\hat S^3$ with eigenvalue $s^3$, $-s\le s^3\le s$.

For half-integer spin a general solution of the system (\ref{sys1}) with
definite sign of the mass term in the second equation possesses definite
sign of $\eta p_0$. This general solution carries a reducible representation
of the improper Poincar\'e group, which is direct sum of two irreps with
opposite signs of $\eta$ and $p_0$. Hence, the general solution contains
$2(2s+1)$ independent components. Since $\eta$ is invariant under the
discrete transformations, the representation carried by the solution remains
reducible under the extended Poincar\'e group.

Thus, for half-integer spin $s$ the sign in the equations (\ref{eqpm}),
(\ref{eqpm1}) coincides with the sign of the product
\be \label{sign}
\eta p_0 S^R_3,
\ee
where the sign of $S^R_3$ distinguishes particles and antiparticles; this
sign is fixed by the choice of the space $V_+$ or $V_-$. In each space the
general solution carries the direct sum of two irreps of the improper
Poincar\'e group characterized by $\sign p_0,\eta$ or $-\sign p_0,-\eta$.
For integer spin in each space the general solution carries the direct sum of
two irreps characterized by fixed intrinsic parity $\eta=1$ and different
signs of $p_0$.

As we saw, the formulation on the base of the set (\ref{31set8}) including
the first order in $\partial/\partial x$ operator $\hat p_\mu\hat\Gamma^\mu$
allows us to fix some characteristics of representations of the
{\it extended} Poincar\'e group. As it was shown in \cite{GitSh00},
for the case of finite-dimensional representations of the Lorentz
group the system (\ref{sys1}) is equivalent to the Bargmann-Wigner equations.
In turn, for half-integer spins the latter equations are equivalent to the
Rarita-Schwinger equations \cite{Ohnuk88}. Hence the above conclusions
concerning the structure of the solutions of the system (\ref{sys1}) are
also valid for mentioned equations.

It is obvious that for the equations fixing not only $m$ and $s$ (like the
Casimir operators (\ref{pw})), but also such characteristics as signs of
the energy or of the charge, only a part of the discrete transformations
forms the symmetry transformations. For example, a discrete symmetry group
of equations (\ref{eqpm}), (\ref{eqpm1}) with definite sign of the mass term
(and therefore discrete symmetry groups of the Dirac and Duffin-Kemmer
equations) includes only the transformations that conserve the sign of
$p_0 S^3_R$.

The transformations $P,C,T_w$ conserve the sign of $p_0 S^3_R$ and
therefore the sign of the mass term in the first order equations under
consideration. Fourth independent transformation (one can take, for example,
inversion $I_x$ or Schwinger time reversal $T_{sch}$) changes the sign of
$p_0 S^3_R$ and correspondingly the sign of the mass term.

The Majorana equations associated with infinite-dimensional irreps of
$SL(2,C)$ \cite{Major32,StoTo68} are more restrictive with respect to
possible symmetry transformations, since allow only the transformations
conserving the sign of $p_0$ \cite{OksTo68,NakGo71}.

On the other hand, a formulation on the base of the set (\ref{31set}) of
commuting operators and the use of representations $(s0)\oplus(0s)$ of the
Lorentz group allows all four independent discrete transformations as symmetry
transformations. To pick out the representation $(s0)\oplus(0s)$ one can use
the Casimir operators of the Lorentz group or (for the subspaces $V_+$ and
$V_-$) the operators $\hat B^3_R$, $\hat S^3_R$; all these operators are
contained in the set (\ref{31set}).
One can show that for subspaces $V_+$ and $V_-$ the system of equations
\be  \label{sys2}
\hat p^2 f(x,\bz)=m^2 f(x,\bz),\quad
\hat S^3_R f(x,\bz)=\pm s f(x,\bz),\quad
i\hat B^3_R f(x,\bz)=\pm s f(x,\bz),
\ee
fixes the spin $s$ \cite{GitSh00}. For definite spin projection $s^3$,
solutions of the system in the rest frame have the form
(in contrast to (\ref{m1}), (\ref{m2}) signs in exponents and in
brackets can be taken independently):
\bea  \label{m3}
&&V_+: \quad e^{\pm imx^0} \left( (z^1)^{s+s^3}(z^2)^{s-s^3}
\pm (\cc\uz_{\dot 1})^{s+s^3}(\cc\uz_{\dot 2})^{s-s^3} \right), \\
&&V_-: \quad e^{\pm imx^0} \left( (\uz^1)^{s+s^3}(\uz^2)^{s-s^3}
\pm (\cc z_{\dot 1})^{s+s^3}(\cc z_{\dot 2})^{s-s^3} \right).
\label{m4}
\eea
The sign in brackets defines the intrinsic parity of the solution.
For half-integer spins the upper sign corresponds to $\eta=1$ and the lower
sign corresponds to $\eta=-1$. For integer spins the upper sign in (\ref{m3})
and the lower sign in (\ref{m4}) correspond to $\eta=1$, and the opposite
signs correspond to $\eta=-1$.
Thus, for each space ($V_+$ or $V_-$) the general solution of the system
has $4(2s+1)$ independent components and carries the reducible
representation of the improper Poincar\'e group. This representation
splits into four irreps labelled by different signs of $\eta$ and $p_0$.

The formulation under consideration allows the coupling of higher spin
with electromagnetic field. This connected with the fact that unlike
the system (\ref{sys1}) the system (\ref{sys2}) contain only one equation
with operator depending on $\partial/\partial x$. (For the system
(\ref{sys1}) the first equation is a consequence of other two only in the
cases $s=1/2$ and $s=1$ \cite{GitSh00}, which correspond to the Dirac
and the Duffin-Kemmer equations.) Particles with definite spin $s$ and mass
$m$ are described by Klein-Gordon equation with polarization,
$$
((\hat p-A(x))^2-\frac e{2s}\hat S^{\mu\nu}F_{\mu\nu}-m^2)\psi(x)=0,
$$
where $\psi(x)$ is transformed under the representation $(s0)\oplus(0s)$
of the Lorentz group \cite{FeyGe58,Iones67,Hurle71,Hurle74,Krugl99}.
For $s=1/2$ this equation is the squared Dirac equation.
Solutions of the Klein-Gordon equation with polarization are casual,
but in contrast to the Dirac and Duffin-Kemmer equations, whose
solutions have $2(2s+1)$ independent components, these solutions
have $4(2s+1)$ independent components (for any sign of energy there are
solutions corresponding to the intrinsic parity $\eta=\pm 1$).

Irrespective of the specific form of relativistic wave equations, the above
analysis shows that in the massive case two of four nontrivial discrete
transformations map any irrep of the improper Poincar\'e group into itself;
these transformations are $P$ and $T_w$.
The operator $P$ labels irreps of the improper group. Wigner time
reversal $T_w$ corresponds to the reversal of the direction of motion
and does not change characteristics of representations of the
Poincar\'e group extended by other discrete transformations ($\eta$ and
signs of energy and $PCT_w$-charge). For example, for spin-1/2 particle at
rest (see (\ref{LRm})) we have
$e^{imx^0}z^\alpha\stackrel{T_w}{\to} e^{imx^0}z_\alpha$, and transformation
reduces to the rotation by the angle $\pi$. In general case $T_w$ is not
reduced to continuos or other discrete transformations. $T_w$ changes
signs both of momentum vector and spin pseudovector, unlike $P$
changing only the sign of vectors.

Two discrete transformations interchange nonequivalent representations of the
improper group extended by the operator $I_z$, which distinguishes particles
and antiparticles. As such transformations one can choose $I_x$ and $I_z$ or
$I_x$ and $C$, in correspondence with that done below, where the first sign
is one of $PCT_w$-charge and the second sign is one of $p_0$:
\bea
\begin{array}{cccccccccc}
& &++& &+-&                 &++& &+-& \\
&C\;\updownarrow &  &\eta=1&  &   \qquad\qquad
                            &  &\eta=-1 &  &\updownarrow\; C \\
& &--& &-+&                 &--& &-+& \\
& &  & \stackrel\longleftrightarrow{I_x} & &
                            & &\stackrel\longleftrightarrow{I_x} &  & \\
 %& &  &\eta=1 &  & & & \eta=-1 & & \\
\end{array}
\eea

A problem of relative parity of particle and antiparticle admits different
treatments. For the first time it was pointed out in \cite{NigFo56}
for spin-1/2 particle; some other cases was studied in
\cite{Silag92,AhlJoG93,Ahluw96}.
Consider the problem in the framework of the representation theory
of the extended Poincar\'e group.

As mentioned above, charge conjugation or $PCT_w$-transformation can't change
the intrisic parity $\eta$, since $C$ and $I_z$ commute with $P$. Therefore,
if one suppose that (i) a particle is described by an irrep of improper
Poincar\'e group and (ii) corresponding antiparticle is described by
$PCT_w$-conjugate (or charge-conjugate) irrep,
then parities of particle and antiparticle must coincide for any spin.
An alternative possibility is instead of (ii) to suppose that antiparticle
is described by the irrep labelled not only by opposite $PCT_w$-charge,
but also by opposite parity $\eta$. In the latter case the irreps describing
particle and antiparticle are not connected by the discrete transformations
$C$ or $PCT_w$.

Usually the relation between parities of particle and antiparticle is
discussed on the base of various wave equations. Consider some relativistic
wave equation describing field with definite spin and mass. As a rule,
a general solution of the equation carries a reducible representation of the
improper Poincar\'e group; irreducible subrepresentations (or their
charge conjugate) are identified with particle and antiparticle fields.
Since different equations have different structure of the solutions,
both possibilities mentioned above can be realized in such an approach.

Consider some examples.
One can suppose that for $s=1/2$ "wave function of antiparticle is a bispinor
charge-conjugate to some "negative-frequency" solution of the Dirac equation"
\cite{BerLiP}. Free Dirac equation have solutions corresponding to two
nonequivalent irreps of the improper Poincar\'e group; these irreps are
characterized by opposite signs of $\eta$ and $p_0$. If
"positive-frequency" solution has intrinsic parity $\eta$, then
"negative-frequency" solution has opposite intrinsic parity $-\eta$, which is
not changed under charge conjugation and intrinsic parities of particle and
antiparticle are opposite. Solutions of the Duffin-Kemmer equation with
different signs of energy have identical intrinsic parity, and similarly we
come to the conclusion that for spin 1 intrinsic parities of particle and
antiparticle are identical. This consideration corresponds to the standard
point of view.
However, the study of a class of equations associated with the
representations $(s0)\oplus(0s)$ of the Lorentz group leads to alternative
conclusion that for integer spin the intrinsic parities of particle and
antiparticle are opposite \cite{AhlJoG93,Ahluw96}.

%%%%%%%%%%%%%%%%%%%%%%%%%%%%%%%%%%%%%%%%%%%%%%%%%%%%%%%%%%%%%%%%%%%%%%%%%%%%%
\section{Group-theoretical derivation of the Dirac equation}

Let us consider a pure group-theoretical derivation of the Dirac equation
in detail.%
\footnote{An heuristic discussion of the problem can be found in
\cite{Ryder88,AhlJoG93,GaiAl95,Ahluw96}.}
The derivation is based on fixing of quantities characterizing the
representations of the extended Poincar\'e group. In addition to the evident
conditions (fixing the mass and spinor representation of the Lorentz group)
it is necessary to demand that states with definite energy possess definite
parity, and also that the states possess definite $PCT_w$-charge.
This formulation shows that the sign of mass term in the Dirac equation
coincides with the sign of the product of three characteristics of the
extended Poincar\'e group representations, namely, the intrinsic
parity, the sign of $PCT_w$-charge, and the sign of energy.
Notice that the consideration and attempts of physical interpretation of
two possible signs of the mass term in the Dirac equation have a long history,
see, in particular, \cite{Marko64,BraLj80,BarZi93,Dvoeg96} and references
therein.

Consider a representation of the extended Poincar\'e group with the
following characteristics:
(i) definite mass $m>0$;
(ii) definite $PCT_w$-charge;
(iii) states with definite sign of energy possess definite intrinsic
parity $\eta=\pm 1$;
suppose then that
(iv) the field $f(x,\bz)$ with above characteristics is linear in $\bz$
(or, that is the same, we fix the representation $(\qq\;0)\oplus(0\;\qq)$
of the spin Lorentz subgroup).

According to (iii), this reducible representation of the extended Poincar\'e
group, which we denote by $T_D$, is the direct sum of two representations
labelled by opposite signs of energy and intrinsic parity.

The conditions (ii) and (iv) restrict our consideration to two functions
\be
f_+(x,\bz)=z\psi_R+\cc\uz\psi_L, \quad f_-(x,\bz)=\uz\psi_R+\cc z\psi_L,
\ee
where we have introduced the columns $\psi_L=(\psi^\da)$,
$\psi_R=(\psi_\alpha)$. These functions correspond to two possible signs of
$PCT_w$-charge. According to (i) there exist functions $f(x,\bz)$ such
that $\hat p_0 f(x,\bz)=p_0 f(x,\bz)$, $\hat p_k f(x,\bz)=0$,
corresponding to particles in the rest frame, where the energy
$p_0=\eps_E m$, $\eps_E=\sign p_0$. According to (iii) these functions
are characterized by the parity $\eta$, defined as an eigenvalue of the
space inversion operator, $Pf(x,\bz)=\eta f(x,\bz)$.
Using the latter equation and the relation (\ref{xz}), we obtain for
functions $f_+(x,\bz)$ and $f_-(x,\bz)$ that
$\psi_R(\oo p)=-\eta\psi_L(\oo p)$ and $\psi_R(\oo p)=\eta\psi_L(\oo p)$
respectively, where $\oo p=(\eps_E m,0)$.
Both the cases can be combined into one equation
\be  \label{R0L0}
\psi_R(\oo p)=\eps_c\eta\psi_L(\oo p),
\ee
where $\eps_c=\sign S^3_R$ is the sign of charge.

The Lorentz transformation of the spinors
$\psi_R(p)=U\psi_R(\oo p)$, $\psi_L(p)=(U^\dagger)^{-1}\psi_L(\oo p)$,
corresponds to the transition to the state characterized by momentum
$P=UP_0U^\dagger$, where $P=p_\mu\sigma^\mu$, $P_0=\eps_E m \sigma_0$,
whence we obtain
\be
{\eps_E m}UU^\dagger = {p_\mu\sigma^\mu}.  \label{uu}
\ee
Taking into account the transformation law of spinors, one can
rewrite the relation (\ref{R0L0}) in the form
$$
\psi_R=\eps_c\eta UU^\dagger\psi_L, \quad
\psi_L=\eps_c\eta (UU^\dagger)^{-1}\psi_R .
$$
Using (\ref{uu}), we express $UU^\dagger$ in terms of momentum,
$$
m\psi_R=\eps_c\eps_E\eta {p_\mu\sigma^\mu}\psi_L, \quad
m\psi_L=\eps_c\eps_E\eta {p_\mu\bar\sigma^\mu}\psi_R,
$$
and combine these two equations into one:
\be
(p_\mu\gamma^\mu-\eps_c\eps_E\eta m)\Psi=0, \quad \gamma^\mu=
\left( \ba{cc} 0 & \sigma^\mu \\ \bar\sigma^\mu & 0 \ea \right), \quad
\Psi=\left( \ba{c} \psi_R \\ \psi_L \ea \right).
\ee
Finally, for the plane waves under consideration one can change the momentum
$p_\mu$ to the corresponding differential operator $\hat p_\mu$.
Since the plane waves form the basis of the representation $T_D$
and the superposition principle is valid for differential equation obtained,
the states belonging to $T_D$ are subjected to the equation
\be
(\hat p_\mu\gamma^\mu-\eps_c\eps_E\eta m)\Psi=0.
\ee

In the above derivation one can use instead of (iii) more restrictive
condition of irreducibility of the representation of the improper Poincar\'e
group. But solutions of the equation obtained will be transformed
under reducible representation obeying the condition (iii) anyway.
The above derivation also shows the impossibility of the derivation of the
Dirac equation only in terms of characteristics of irreps of the proper or
improper Poincar\'e group, since the Dirac equation connects signs
of the energy $\eps_E$, of the parity $\eta$, and of the charge $\eps_c$,
which characterize representations of the extended Poincar\'e group.

%%%%%%%%%%%%%%%%%%%%%%%%%%%%%%%%%%%%%%%%%%%%%%%%%%%%%%%%%%%%%%%%%%%%%%%%%%%%%
\section{Discrete symmetries of the relativistic wave equations.
         Massless case}

For standard massless fields with discrete spin the eigenvalues of the
Casimir operators $\hat p^2$ and $\hat W^2$ are equal to zero (see, e.g.,
\cite{Tung85}). As a consequence such massless fields obey the conditions
\be
\hat W_\mu f(x,\bz) = \lambda \hat p_\mu f(x,\bz),
\ee
where $\lambda$ is the helicity. For the component $\hat W_0$ we have
\be  \label{m0}
\hat{\bf p}\hat{\bf S}f(x,\bz)=\lambda \hat p_0 f(x,\bz).
\ee
The transformations $P$ and $C$ change the sign in equation (\ref{m0}); on
the other hand, the transformations $I_x$, $I_z$, $T_{sch}$, which change the
sign of mass term of the Dirac equation, are symmetry transformations
of equation (\ref{m0}). Discrete symmetries of equation (\ref{m0}) are
generated by three independent transformations, for example, $I_x$, $CP$,
$T_w$, the first of which is not a symmetry transformation of the Dirac
equation.

The Weyl equation $\hat {\bf p}{\bf\sigma}\Psi(x)=\pm \hat p^0 \Psi(x)$
is the particular case of (\ref{m0}) corresponding to helicity $\pm 1/2$;
they can be obtained by the substitution of the function
$f(x,z)=\Psi_\alpha(x)z^\alpha$ into (\ref{m0}).

Massless irreps of the proper Poincar\'e group (see Table 4) are labelled
by two numbers, namely, by the helicity $\lambda={\bf pS}/p_0$ and by the
sign of $p_0$. If we not claim that the states possess definite parity, then
instead of the subspaces  $V_+$ and $V_-$ it is natural to consider four
subspaces of functions $f(x,z)$, $f(x,\uz)$, $f(x,\cc z)$, $f(x,\cc\uz)$.

For definite chirality $s$ in any subspace equation (\ref{m0}) has four
solutions which correspond to the motion along the axis $x^3$. This solutions
are labelled by the signs of helicity and $p_0$. Considering the action of
the operators $C$ and $I_z$, it is easy to see that these functions describe
particles which is not coincide with their antiparticles.

For $p_0>0$ we have for particles
\bea
&&\lambda=s:\qquad\! e^{i(px^0+px^3)} (z^1)^{2s}, \quad
e^{i(px^0+px^3)} (\cc\uz_{\dot 1})^{2s}, \quad        \\
&&\lambda=-s:\quad e^{i(px^0+px^3)} (z^2)^{2s},
\quad e^{i(px^0+px^3)} (\cc\uz_{\dot 2})^{2s},
\eea
and for antiparticles
\bea
&&\lambda=s:\qquad\! e^{i(px^0+px^3)} (\uz^1)^{2s}, \quad
e^{i(px^0+px^3)} (\cc z_{\dot 1})^{2s}, \quad         \\
&&\lambda=-s:\quad e^{i(px^0+px^3)} (\uz^2)^{2s},\quad
e^{i(px^0+px^3)} (\cc z_{\dot 2})^{2s}.
\eea
The operators $P$ and $C$ interchange the states with opposite chirality.
The operator $I_z$, interchanging the states with opposite $PCT_w$-charge,
conserves the signs of the chirality and of the energy.

The signs of the helicity and of the chirality are changed simultaneously
under the discrete transformations. But unlike the parity $\eta$, which also is
not changed under the discrete transformations, the sign connecting the helicity
and the chirality characterizes equivalent representations of the extended
Poincar\'e group.

Above we have developed the description of particles which differ from
their antiparticles. As an example let us consider now the description of
pure neutral massless spin-1 particle (photon) in terms of field on the
Poincar\'e group. Such a particle is its own antiparticle (i.e. it has zero
$PCT_w$-charge) and possesses chirality $\pm 1$.
For the quadratic in $\bz=(z,\uz,\cc z,\cc\uz)$ functions on the group
these conditions are satisfied only by the fields depending on
$z^\alpha\uz^\beta$, $\cc z_\da \cc\uz_\db$ and being eigenfunctions of
$\hat S^3_R$ with zero eigenvalue.

Thus, a pure neutral massless spin-1 particle should be described by the
function
\be \label{m00}
f(x,\bz)= \chi_{\alpha\beta}(x) z^\alpha\uz^\beta +
            \psi^{\da\db}(x)     \cc z_\da \cc\uz_\db =
\qq F_{\mu\nu}(x) q^{\mu\nu} ,
\ee
where
\bea \label{31qq}
&&q_{\mu\nu}=-q_{\nu\mu}=\qq \left((\sigma_{\mu\nu})_{\alpha\beta}z^\alpha
  \uz^\beta + (\bar\sigma_{\mu\nu})_{\da\db} \cc z^{\da}\cc\uz^{\db}\right),
  \quad \cc q_{\mu\nu}= q_{\mu\nu},
\\   \label{Fmn}
&&F_{\mu\nu}(x)=-2\left( (\sigma_{\mu\nu})_{\alpha\beta}\chi^{\alpha\beta}(x)+
  (\bar\sigma_{\mu\nu})_{\da\db}\psi^{\da\db}(x)\right).
\eea
The functions $\chi_{\alpha\beta}(x)$ and $\psi_{\da\db}(x)$ must be
symmetric in their indices; in opposite case by virtue of the constraint
$z^1\uz^2-z^2\uz^1=1$ (which is a consequence of unimodularity of $SL(2,C)$)
the field (\ref{m00}) will contain components $\chi_{[\alpha\beta]}(x)$ and
$\psi_{[\da\db]}(x)$ of zero spin. Therefore the formulations in terms of
$\chi_{\alpha\beta}(x)$, $\psi_{\da\db}(x)$ and $F_{\mu\nu}(x)$ are
equivalent. Left and right fields can be described by
\bea
&&F^L_{\mu\nu}(x)=F_{\mu\nu}(x)-i\tilde F_{\mu\nu}(x) =
-4(\bar\sigma_{\mu\nu})_{\da\db}\psi^{\da\db}(x),
\label{FL} \\ \label{FR}
&&F^R_{\mu\nu}(x)=F_{\mu\nu}(x)+i\tilde F_{\mu\nu}(x) =
-4(\sigma_{\mu\nu})_{\alpha\beta}\chi^{\alpha\beta}(x),
\eea
where $\tilde F_{\mu\nu}(x)=\qq\epsilon_{\mu\nu\rho\sigma}F^{\rho\sigma}$.

To describe the states with definite helicity, the function (\ref{m00})
should obey equation (\ref{m0}) for $\lambda=\pm 1$,
\be  \label{m00e}
(\hat{\bf p}\hat{\bf S}\mp \hat p^0)f(x,\bz)=0.
\ee
For $p_0>0$ equation (\ref{m00e}) has four solutions which correspond to
the motion along the axis $x^3$. These solutions are distinguished by signs
of helicity $\lambda$ and chirality:
\bea
&&\lambda=1:\qquad\! e^{i(px^0+px^3)} z^1\uz^1, \quad
e^{i(px^0+px^3)} \cc z_{\dot 1}\cc\uz_{\dot 1}, \quad   \\
&&\lambda=-1:\quad e^{i(px^0+px^3)} z^2\uz^2,
\quad e^{i(px^0+px^3)} \cc z_{\dot 2}\cc\uz_{\dot 2}.
\eea
Fixing the sign connecting helicity and chirality (this sign distinguishes
the equivalent representations of the Poincar\'e group), we obtain two
solutions corresponding to two polarization states.

Substituting the functions
$f_L(x,\bz)=\psi^{\da\db}(x)\cc z_\da \cc\uz_\db$ and
$f_R(x,\bz)=\chi_{\alpha\beta}(x) z^\alpha\uz^\beta$ into (\ref{m00e})
(for $\lambda=\pm 1$ respectively) and in accordance with (\ref{FL})
(\ref{FR}) going over to the vector notation, we obtain equations for
$F^L_{\mu\nu}(x)$ and $F^R_{\mu\nu}(x)$,
\be
\partial^\mu F^L_{\mu\nu}(x)=0, \quad \partial^\mu F^R_{\mu\nu}(x)=0;
\ee
then, constructing their linear combinations, we come to the Maxwell
equations:
\be \label{Mw}
\partial^\mu F_{\mu\nu}(x)=0, \quad \partial^\mu \tilde F_{\mu\nu}(x)=0.
\ee
If we introduce complex potentials $A_\mu$,
$F_{\mu\nu}(x)=\partial_\mu A_\nu - \partial_\nu A_\mu$, then the second
equation is satisfied identically.

Taking into account the action of operators of the discrete transformations on
$\bz$ (see (\ref{xz})), we find for $q^{\mu\nu}$:
$q^{\mu\nu}\stackrel{P}{\to} (-1)^{\delta_{0\mu}+ \delta_{0\mu}}q^{\mu\nu}$,
$q^{\mu\nu}\stackrel{I_z}{\to} -q^{\mu\nu}$,
$q^{\mu\nu}\stackrel{C}{\to} q^{\mu\nu}$, whence as a consequence of
(\ref{m00}) we obtain
\bea
&&P:\quad F_{\mu\nu}(x)\to (-1)^{\delta_{0\mu}+\delta_{0\mu}}F_{\mu\nu}(\bar x),
\quad A_\mu(x) \to -(-1)^{\delta_{0\mu}} A_\mu(\bar x); \\
&&I_z:\quad F_{\mu\nu}(x)\to - F_{\mu\nu}(x), \quad A_\mu(x)\to - A_\mu(x), \\
&&C:\quad F_{\mu\nu}(x)\to \cc F_{\mu\nu}(x), \quad A_\mu(x)\to \cc A_\mu(x).
\eea
It is easy to see that the charge conjugation transformation, acting on the
function (\ref{m00}) as complex conjugation, interchanges states with
opposite helicities and thus can't be considered for left and right fields
separately (as it was pointed out, e.g., in \cite{Ohnuk88}). The
transformation $I_3$ for functions (\ref{m00}) is the identity transformation.

Unlike the initial equations (\ref{m00e}), where the sign at $p_0$ is
changed under space reflection and charge conjugation, equations (\ref{Mw})
are invariant under these transformations, since the left and right fields
are contained in $F_{\mu\nu}(x)$ on an equal footing. Thus four discrete
transformations $P,I_x,C,T_w$ are symmetry transformations of equations
(\ref{Mw}).

Formally one can define two real fields, $F^{(1)}_{\mu\nu}(x)$ and
$F^{(2)}_{\mu\nu}(x)$, as real and imaginary parts of $F_{\mu\nu}(x)$,
which satisfy the same equations (\ref{Mw}) and are characterized by
opposite parities with respect to charge conjugation $C$.
However, the real field $F^{(i)}_{\mu\nu}(x)$ can't describe the states with
definite helicity, since according to (\ref{Fmn}) includes both left and
right components.
It is necessary also to note that $F^L_{\mu\nu}(x)$ and $F^R_{\mu\nu}(x)$
are no classical electromagnetic fields themselves.
 %but are complex quantities quite different conceptually.
They can be treated as wave functions of left-handed and right-handed photons
\cite{Ohnuk88,AkhBe81,Bialy94}.
This shows that, as for all other cases considered above, the explicit
realization of the representations of the Poincar\'e group in the space of
functions on the group corresponds to the construction of one-particle
sector of the quantum field theory.

%%%%%%%%%%%%%%%%%%%%%%%%%%%%%%%%%%%%%%%%%%%%%%%%%%%%%%%%%%%%%%%%%%%%%%%%%%%%%%
\section{Conclusion}

We have shown that the representation theory of the proper Poincar\'e group
implies the existence of five nontrivial independent discrete transformations
corresponding to involutory automorphisms of the group.
As such transformations one can choose  space reflection $P$,
inversion $I_x$, charge conjugation $C$, Wigner time reversal $T_w$;
the fifth transformation for most fields of physical interest (except
the Majorana field) is reduced to the multiplication by a phase factor.

Considering discrete automorphisms as operators acting in the space of the
functions on the Poincar\'e group, we have obtained the explicit form for the
discrete transformations of arbitrary spin fields without the use of any
relativistic wave equations or special assumptions. The examination of the
action of automorphisms on the operators, in particular, on the generators of
the Poincar\'e group, ensures the possibility to get transformation laws of
corresponding physical quantities. The analysis of the scalar field on the
group allows us to construct explicitly the states corresponding to
representations of the extended Poincar\'e group, and also to give the
classification of the solutions of various types of relativistic wave
equations with respect to representations of the extended group.

Since in the general case a relativistic wave equation can fix some
characteristics of the extended Poincar\'e group representation, which are
changed under the discrete transformations, only a part of the discrete
transformations forms symmetry transformations of the equation.
In particular, discrete symmetries of the Dirac equation and of the Weyl
equation are generated by two different sets of the discrete transformations
operators, $P,C,T_w$ and $PC,I_x,T_w$ respectively.

Being based on the concept of the field on the group and on the consideration
of the group automorphisms, the approach developed can be applied to the
analysis of discrete symmetries in other dimensions and also to other
spacetime symmetry groups.

%%%%%%%%%%%%%%%%%%%%%%%%%%%%%%%%%%%%%%%%%%%%%%%%%%%%%%%%%%%%%%%%%%%%%%%%%%%%%
\section*{Acknowledgments}

The authors would like to thank A.Grishkov for useful discussions.
I.L.B. and A.L.Sh. are grateful to the Institute of Physics at the
University of S{\~a}o Paulo for hospitality.
This work was partially supported by Brazilian Agencies CNPq (D.M.G.) and
\mbox{FAPESP} (I.L.B., D.M.G. and A.L.Sh.).

%%%%%%%%%%%%%%%%%%%%%%%%%%%%%%%%%%%%%%%%%%%%%%%%%%%%%%%%%%%%%%%%%%%%%%%%%%%%%
\appendix
\section{Operators in the space of functions on the Poincar\'e group}

1. Left and right generators of $SL(2,C)$ in the space of the functions
on the group.

The left and right spin operators are given in the form \cite{GitSh00}
\bea
&&\hat S_k=\frac 12 (z\sigma_k\partial _z -\cc z\cc\sigma_k\partial _{\ccc z}\,)+... \; ,
\nonumber \\ \label{SL}
&&\hat B_k=
\frac i2 (z\sigma_k\partial _z + \cc z\cc\sigma_k\partial _{\ccc z}\,)+... \; ,
\quad z=(z^1\; z^2), \quad
\partial_z=(\partial/\partial{z^1}\; \partial/\partial{z^2})^T ;
\\
&&\hat S_k^R=-\frac 12 (\chi\cc\sigma_k\partial_\chi
-\cc\chi\sigma_k\partial _{\ccc \chi}\,)+... \; ,
\nonumber \\  \label{SR}
&&\hat B_k^R=
-\frac i2 (\chi\cc\sigma_k\partial_\chi + \cc\chi\sigma_k\partial _{\ccc \chi}\,)+... \; ,
\quad \chi=(z^1\;\uz^1), \quad
\partial_\chi=(\partial/\partial{z^1}\; \partial/\partial{\uz^1})^T ;
\eea
Dots in the formulae mean analogous expressions obtained by the substitutions
$z\to \uz=(\uz^1\;\uz^2)$, $\chi\to \chi'=(z^2\; \uz^2)$.
One can rewrite two first formulae in four-dimensional notation:
\be  \label{31S}
\hat S^{\mu\nu}=
 \qq ( (\sigma^{\mu\nu})^{\;\;\beta}_\alpha z^\alpha \partial_\beta +
 (\bar\sigma^{\mu\nu})^\da_{\;\;\db} \cc\uz_\da \underline\partial^\db )-c.c.,
\ee
where $\partial_{\alpha}=\partial/\partial z^{\alpha}$,
$\underline\partial^{\da}=\partial/\partial \cc\uz_{\da}$,
\be \label{sigmn}
(\sigma^{\mu\nu})_\alpha^{\;\;\beta} = -\frac i4
(\sigma^\mu\bar\sigma^\nu-\sigma^\nu\bar\sigma^\mu)_\alpha^{\;\;\beta},\quad
(\bar\sigma^{\mu\nu})^\da_{\;\;\db}= -\frac i4
(\bar\sigma^\mu\sigma^\nu-\bar\sigma^\nu\sigma^\mu)^\da_{\;\;\db},
\ee
and $c.c.$ is complex conjugate term corresponding to the action in the space
of polynomials in $\uz^\alpha,\cc z_{\dot\alpha}$.

2. Operators $\hat\Gamma^\mu$ and equations for definite mass and spin.

The equations for functions on the Poincar\'e group
\bea
&& \hat p^2 f(x,z,\cc\uz) = m^2 f(x,z,\cc\uz),           \label{fmeq0}
\\
&&\hat p_\mu \hat\Gamma^\mu f(x,z,\cc\uz) = ms f(x,z,\cc\uz), \label{fseq0}
\eea
where
\be  \label{31Gmu}
\hat\Gamma^\mu= \frac 12 \left(
 \bar\sigma^{\mu\da\alpha}\cc\uz_{\da}\partial_{\alpha}+
 {\sigma^\mu}_{\alpha\da}z^{\alpha}\underline\partial^{\da} \right)-c.c.,
\ee
describe a particle with fixed mass $m>0$ and spin $s$, if we suppose that
$f(x,z,\cc\uz)$ is a polynomial of the power $2s$ in $z,\cc\uz$
\cite{GitSh00}. Analogous statement also holds for polynomial in $\uz,\cc z$
functions $f(x,\uz,\cc z)$.
Operators $\hat\Gamma^\mu$ and $\hat S^{\mu\nu}$ obey the commutation
relations
\be
[\hat S^{\lambda\mu},\hat\Gamma^\nu]
=i(\eta^{\mu\nu}\hat \Gamma^\lambda - \eta^{\lambda\nu}\hat \Gamma^\mu ),
\quad
[\hat\Gamma^\mu,\hat\Gamma^\nu]=-i\hat S^{\mu\nu}, \label{comg}
\ee
of the group $SO(3,2)$, which coincide with the commutation relations of
matrices $\gamma^\mu /2$. These operators together with the chirality
operator
\be  \label{31G5}
\hat\Gamma^5 = \qq \left( z^{\alpha} \partial_\alpha
 -\cc\uz_{\dot\alpha} \underline\partial^{\da} \right)-c.c.,
\ee
and the operators $\uGamma^\mu=i[\hat\Gamma^\mu,\hat\Gamma^5]$, $\hat S^R_3$
form the set of 16 operators, which conserve the power of polinomials
$f(x,z,\cc\uz)$ in $z,\cc\uz$.

Going over to spin-tensor notation, one can find that equation (\ref{fseq0})
for $s=1/2$ is transformed to the Dirac equation and for $s=1$ is
transformed to the Duffin-Kemmer equation. In general case, going over to
spin-tensor notation, we obtain that the system (\ref{fmeq0})-(\ref{fseq0})
for polynomial of power $2s$ in $z,\cc\uz$ functions $f(x,z,\cc\uz)$ is
transformed to the system of the Klein-Gordon equation and symmetric Bhabha
equation \cite{GitSh00}. The latter system is equivalent to the
Bargmann-Wigner equations \cite{LoiOtS97,GitSh00}.

%%%%%%%%%%%%%%%%%%%%%%%%%%%%%%%%%%%%%%%%%%%%%%%%%%%%%%%%%%%%%%%%%%%%%%%%%%%%%%


\begin{thebibliography}{10}

\bibitem{Wig39}
E.P. Wigner, ``On unitary representations of the inhomogeneous {L}orentz
  group,'' {\em Ann. Math.} {\bf 40}, 149--204 (1939).

\bibitem{Macke68}
G.~Mackey, {\em Induced Representations of Groups and Quantum Mechanics}
  (Benjamin, New York, 1968).

\bibitem{BarRa77}
A.O. Barut and R.~Raczka, {\em Theory of Group Representations and
  Applications} (PWN, Warszawa, 1977).

\bibitem{Tung85}
{W.-K.} Tung, {\em Group Theory in Physics} (World Scientific, Singapore,
  1985).

\bibitem{KimNo86}
Y.S. Kim and M.E. Noz, {\em Theory and Applications of the {P}oincar\'e Group}
  (Reidel, Dordrecht, 1986).

\bibitem{Ohnuk88}
Y.~Ohnuki, {\em Unitary Representations of the {P}oincar\'e Group and
  Relativistic Wave Equations} (World Scientific, Singapore, 1988).

\bibitem{GelMiS63}
I.M. Gel'fand, R.A. Minlos, and Z.Ya. Shapiro, {\em Representations of the
  Rotation and {L}orentz Groups and their Applications} (Pergamon press,
  Oxford, 1963).

\bibitem{Wigne32}
E.P. Wigner, ``{\"{U}}ber die operation der zeitumkehr in der quantenmechanik,''
  {\em Nachr. Ges. Wiss. G\"{o}tt.}, 546--559 (1932).

\bibitem{Schwi51}
J.~Schwinger, ``The theory of quantized fields. {I},'' {\em Phys. Rev.} {\bf
  82}, 914--927 (1951).

\bibitem{UmeKaT54}
H.~Umezava, S.~Kamefuchi, and S.~Tanaka, ``On the time reversal in the
  quantized field theory,'' {\em Prog. Theor. Phys.} {\bf 12}, 383--400 (1954).

\bibitem{LeeWi66}
T.D. Lee and G.C. Wick, ``Space inversion, time reversal, and other discrete
  symmetries in local field theories,'' {\em Phys. Rev.} {\bf 148}, 1385--1404
  (1966).

\bibitem{BenTu81}
I.M. Benn and R.W. Tucker, ``Extended {L}orentz invariance and field theory,''
  {\em J. Phys. A} {\bf 14}, 1745--1759 (1981).

\bibitem{Shiro58}
{Yu}.M. Shirokov, ``A group-theoretical consideration of the basis of
  relativistic quantum mechanics. {IV}. {S}pace reflections in quantum
  theory,'' {\em Sov. Phys. -- JETP} {\bf 7}, 493--498 (1958).

\bibitem{Shiro60}
{Yu}.~M. Shirokov, ``Space and time reflections in relativistic theory,'' {\em
  Nucl. Phys.} {\bf 15}, 1--12 (1960).

\bibitem{Wigne64}
E.P. Wigner, ``Unitary representations of the inhomogeneous {L}orentz group
  including reflections,'' in F.~G\"{u}rsey, editor, {\em Group Theoretical
  Concepts and Methods in Elementary Particle Physics}, pages 37--80, (Gordon
  and Breach, New York, 1964).

\bibitem{Miche64}
L.~Michel, ``Invariance in quantum mechanics and group extension,'' in
  F.~G\"{u}rsey, editor, {\em Group Theoretical Concepts and Methods in
  Elementary Particle Physics}, pages 135--200, (Gordon and Breach,
  New York, 1964).

\bibitem{Kuo71a}
Kuo T.K., ``Internal-symmetry groups and their automorphisms,'' {\em Phys. Rev.
  D} {\bf 4}, 3620--3637 (1971).

\bibitem{Silag92}
Z.K. Silagadze, ``On the internal parity of antiparticles,'' {\em Sov. J. Nucl.
  Phys.} {\bf 55}, 392--396 (1992).

\bibitem{GitSh00}
D.M. Gitman and A.L. Shelepin.
\newblock ``Fields on the {P}oincar\'e group: Arbitrary spin description and
  relativistic wave equations,'' hep-th/0003146.

\bibitem{Marko64}
M.A. Markov, {\em Neutrino} (Nauka, Moscow, 1964).

\bibitem{BraLj80}
J.~Brana and K.~Ljolje, ``Dual symmetry and the {D}irac field theory,'' {\em
  Fizika} {\bf 12}, 287--319 (1980).

\bibitem{BarZi93}
A.O. Barut and G.~Ziino, ``On parity conservation and the question of the
  'missing' (right-handed) neutrino,'' {\em Mod. Phys. Lett. A} {\bf 8},
  1011--1020 (1993).

\bibitem{Ahluw96}
Ahluwalia D.V., ``Theory of neutral particles: {M}c{L}ennan-{C}ase construct
  for neutrino, its generalization, and new wave equation,'' {\em Int. J. Mod.
  Phys. A} {\bf 11}, 1855--1874 (1996).

\bibitem{Dvoeg96}
V.V. Dvoeglazov, ``Extra {D}irac equations,'' {\em Nuovo Cimento B} {\bf 111},
  483--496 (1996).

\bibitem{StrWi64}
R.F. Streater and A.S. Wightman, {\em {PCT}, Spin and Statistics, and all that}
  (Reading, Massachusetts, 1964).

\bibitem{BucKu95}
I.L.Buchbinder and S.M.Kuzenko, {\em Ideas and Methods of Supersymmetry and
  Supergravity} (IOP Publishing Ltd, Bristol, 1995).

\bibitem{Vilen68t}
N.Ya. Vilenkin, {\em Special Functions and the Theory of Group Representations}
  (AMS, Providence, 1968).

\bibitem{ZhelSc83}
D.P. Zhelobenko and A.I. Schtern, {\em Representations of {L}ie Groups} (Nauka,
  Moscow, 1983).

\bibitem{Sachs87}
R.G. Sachs, {\em The Physics of Time Reversal} (The University of Chicago
  Press, Chicago, 1987).

\bibitem{KayGo83}
B.~Kayser and A.S. Goldhaber, ``{CPT} and {CP} properties of {M}ajorana
  particles, and the consequences,'' {\em Phys. Rev. D} {\bf 28}, 2341--2344
  (1983).

\bibitem{Wigne59}
E.P. Wigner, {\em Group Theory and its Application to the Quantum Mechanics of
  Atomic Spectra} (Academic Press, New York, 1959).

\bibitem{LanLi3}
L.D. Landau and E.M. Lifschitz, {\em Quantum Mechanics}, volume 3 of Course of
  Theoretical Physics (Pergamon, Oxford, 1977).

\bibitem{BieLo81}
L.S. Biedenharn and J.D. Louck, {\em Angular Momentum in Quantum Physics}
  (Addison-Wesley, Reading, Massachusetts, 1981).

\bibitem{ShaLe74}
R.~Shaw and J.~Lever, ``Irreductible multiplier corepresentations of the
  extended {P}oincar\'e group,'' {\em Commun. Math. Phys.} {\bf 38}, 279--297
  (1974).

\bibitem{GavGi00}
S.P. Gavrilov and D.M. Gitman.
\newblock ``Quantization of point-like particles and consistent relativistic
  quantum mechanics,'' hep-th/0003112.

\bibitem{Hai69}
N.X. Hai, ``Harmonic analysis on the {P}oincar\'e group, {I}. {G}eneralized
  matrix elements,'' {\em Commun. Math. Phys.} {\bf 12}, 331--350 (1969).

\bibitem{Major32}
E.~Majorana, ``Teoria relativistica di particelle con momento intrinseco
  arbitrario,'' {\em Nuovo Cimento} {\bf 9}, 335--344 (1932).

\bibitem{StoTo68}
D.{Tz}. Stoyanov and I.T. Todorov, ``{M}ajorana representations of the
  {L}orentz group and infinite-component fields,'' {\em J. Math. Phys.} {\bf
  9}, 2146--2167 (1968).

\bibitem{OksTo68}
A.I. Oksak and I.T. Todorov, ``Invalidity of {TCP}-theorem for
  infinite-component fields,'' {\em Commun. Math. Phys.} {\bf 11}, 125--130
  (1968).

\bibitem{NakGo71}
S.~Naka and T.~Got\={o}, ``{$C,P$} and {$T$} in {$c$}-number infinite component
  wave function,'' {\em Prog. Theor. Phys.} {\bf 45}, 1979--1986 (1971).

\bibitem{FeyGe58}
R.P. Feynman and M.~Gell-Mann, ``Theory of {F}ermi interaction,'' {\em Phys.
  Rev.} {\bf 109}, 193--198 (1958).

\bibitem{Iones67}
N.J. Ionesco-Pallas, ``Relativistic {S}chr\"{o}dinger equation for a particle
  with arbitrary spin,'' {\em J. of the Franklin Inst.} {\bf 284}, 243--250
  (1967).

\bibitem{Hurle71}
W.J. Hurley, ``Relativistic wave equations for particles with arbitrary spin,''
  {\em Phys. Rev. D} {\bf 4}, 3605--3616 (1971).

\bibitem{Hurle74}
W.J. Hurley, ``Invariant bilinear forms and the discrete symmetries for
  relativistic arbitrary-spin fields,'' {\em Phys. Rev. D} {\bf 10}, 1185--1200
  (1974).

\bibitem{Krugl99}
S.I. Kruglov.
\newblock ``Pair production and solutions of the wave equation for particles
  with arbitrary spin,'' hep-ph/9908410.

\bibitem{NigFo56}
Nigam B.P. and Foldy L.L., ``Representation of charge conjugation for {D}irac
  fields,'' {\em Phys. Rev.} {\bf 102}, 1410--1412 (1956).

\bibitem{AhlJoG93}
Ahluwalia D.V., Jonson M.B., and Goldman T., ``A
  {B}argmann-{W}ightman-{W}igner-type quantum field theory,'' {\em Phys. Lett.
  B} {\bf 316}, 102--108 (1993).

\bibitem{BerLiP}
V.B. Berestetskii, E.M. Lifshitz, and L.P. Pitaevskii, {\em Relativistic
  Quantum Theory} (Pergamon, New York, 1971).

\bibitem{Ryder88}
L.~Ryder, {\em Quantum Field Theory} (Cambrige University, Cambrige, 1988).

\bibitem{GaiAl95}
F.H. Gaioli and E.T.G. Alvarez, ``Some remarks about intrinsic parity in
  {R}yder's derivation of the {D}irac equation,'' {\em Am. J. Phys.} {\bf 63},
  177--178 (1995).

\bibitem{AkhBe81}
A.I. Akhiezer and V.B. Berestetskii, {\em Quantum Electrodynamics} (Nauka,
  Moscow, 1981).

\bibitem{Bialy94}
I.~Bia{\l}ynicki-Birula, ``On the wave function of the photon,'' {\em Acta
  Phys. Pol., Ser. B} {\bf 86}, 97--116 (1994).

\bibitem{LoiOtS97}
R.-K. Loide, I.~Ots, and R.~Saar, ``{B}habha relativistic wave equations,''
  {\em J. Phys. A} {\bf 30}, 4005--4017 (1997).

\end{thebibliography}
\end{document}